\documentclass[preprint,3p,10pt]{elsarticle}
\usepackage{lineno,hyperref}
\usepackage{amsmath,bm,geometry}
\usepackage{amsfonts}
\usepackage{color}
\usepackage{nomencl,etoolbox,ragged2e,siunitx}
\usepackage{mathtools}
\usepackage{multirow}
\usepackage{caption}
\usepackage{parskip}
\usepackage{calligra}
\usepackage{makeidx}
\usepackage{tikz}
\usepackage{mathrsfs}
\usepackage{bbm}
\usepackage{algorithm}
\usepackage{algorithmic}
\usepackage{amsthm}
\usepackage{graphicx}
\usepackage{subfigure}
\usepackage{hyperref}
\hypersetup{pdfauthor=author}
\usetikzlibrary{matrix}
\makeindex
\DeclareMathAlphabet{\mathcalligra}{T1}{calligra}{m}{n}
\begin{document}
\title{A unified gas-kinetic particle method for frequency-dependent radiative transfer equations with isotropic scattering process on unstructured mesh}
\author[ad1]{Yuan Hu}
\ead{hu_yuan@iapcm.ac.cn}
\author[ad1]{Chang Liu\corref{cor1}}
\ead{liuchang@iapcm.ac.cn}
\address[ad1]{Institute of Applied Physics and Computational Mathematics, Beijing, China}
\cortext[cor1]{Corresponding author}

\begin{abstract}
In this paper, we extend the unified kinetic particle (UGKP) method to 
the frequency-dependent radiative transfer equation with both absorption-emission and scattering processes.
The extended UGKP method could not only capture the diffusion and free transport limit, 
but also provide a smooth transition in the physical and frequency space in the regime between the above two limits. 
The proposed scheme has the properties of asymptotic-preserving, regime-adaptive, and entropy-preserving,
which make it an accurate and efficient scheme in the simulation of multiscale photon transport problems.
The methodology of scheme construction is a coupled evolution of macroscopic energy equation and the microscopic radiant intensity equation,
where the numerical flux in macroscopic energy equation and the closure in microscopic radiant intensity equation are constructed based on the integral solution.
Both numerical dissipation and computational complexity are well controlled especially in the optical thick regime.
A 2D multi-thread code on a general unstructured mesh has been developed.
Several numerical tests have been simulated to verify the numerical scheme and code, covering a wide range of flow regimes.
The numerical scheme and code that we developed are highly demanded and widely applicable in the high energy density engineering applications. 
\end{abstract}
\maketitle

\section{Introduction}\label{Introduction}
The thermal radiative transfer (TRT) equations, 
which describe the time evolution of radiative intensity and its interaction with the background material, 
has wide applications in astrophysics, atmospheric physics, 
inertial confinement fusion (ICF), high temperature flow systems, plasma physics, etc \cite{lan2022dream,chen2022determination}. 
It comprises the kinetic radiation transport equation which describes the photon transport in the background material 
and the material energy equation which describes the energy exchange between radiation and background material. 
These two equations are coupled by the absorption-emission process that is characterized by the material opacity.
The nonlinear dependency of the material opacity and material temperature makes the system difficult to solve \cite{wollaber2016four,noebauer2019monte}. 
In addition, the high-dimensionality of the equation greatly increases the computational cost.
To develop numerical methods with high accuracy and high efficiency has became an important topic for the past decades.

Generally, the numerical methods for radiative transfer equations can be categorized into the deterministic method and the stochastic method. 
The deterministic methods include the macroscopic moment methods 
\cite{frank2006partial,carrillo2008numerical,vikas2013radiation,alldredge2016approximating} 
and microscopic discrete ordinate SN method 
\cite{hunter2013comparison,coelho2014advances,chen2015chebyshev,roos2016conservation}.
The moment methods propose a closure to the radiant intensity by expanding it in a specific functional space 
\cite{fu2022asymptotic}.
The SN methods directly discretize the velocity space using a specific quadrature.
For stochastic methods, the most commonly used Monte Carlo (MC) method 
\cite{fleck1971implicit,lucy1999computing,hayakawa2007coupled} exploits random numbers to simulate the interactions of individual radiation particles with the background material. 
The MC method is more efficient in optically thin regimes especially for the multidimensional cases, 
and does not suffer from the ray effect compared with the deterministic method.
The implicit Monte Carlo (IMC) method proposed by Fleck and Cummings 
\cite{fleck1971implicit}
is a popular Monte Carlo method for solving the TRT equations. 
This method approximates the rapid, dynamic timescale of photon absorption-emission processes via effective scattering events by Fleck factor, according to which the nonlinear TRT equations are reformulated into a system of linearized equations and solved by the standard Monte Carlo method. 
However, it is generally noticed that the IMC method becomes inefficient in optically thick region 
when the photon mean free path is much smaller than the flow characteristic length, and the particle collision becomes dominated.
In such regime, a great number of effective scattering events are calculated during a time step, 
which significantly increases the computational cost.
Efforts have been made to improve the efficiency of the IMC method in optically thick regions 
\cite{fleck1984random,giorla1987random}, 
such as the implicit Monte Carlo diffusion (IMD) \cite{gentile2001implicit}, 
discrete diffusion Monte Carlo (DDMC) \cite{densmore2007hybrid,densmore2012hybrid} methods, 
as well as the moment-based scale-bridging method 
\cite{park2012consistent,park2013efficient,densmore2015monte,hammer2019multi,park2019multigroup}. 
The IMD and DDMC methods are transport-diffusion hybrid methods which simulate the TRT equations with diffusion approximation in optically thick regions and the standard IMC method in other regions. 
For the transport-diffusion hybrid method, special efforts need to be made for the domain decomposition and the information exchange at transport-diffusion interfaces. 
For the moment-based scale-bridging method, a coupled high-order and low-order (HO-LO) equations are solved to improve the overall efficiency of simulation for the TRT equations. 
Numerical principles such as the asymptotic-preserving property, the regime-adaptive property, and entropy-preserving property have been proposed that provide guidance for the multiscale numerical scheme construction \cite{jin1999ap,jin2010asymptotic,guo2023unified}.

The unified gas kinetic scheme (UGKS) has been constructed first to simulate both continuum and rarefied flow \cite{xu2010unified}, 
and then extended for the TRT problems 
\cite{mieussens2013asymptotic,sun2015asymptoticgray,sun2015asymptoticfrequency,sun2017implicit,sun2017multidimensional,sun2018asymptotic}. 
The UGKS utilizes a finite volume formulation to solve the macroscopic transport and material energy equations, 
where the DOM method is employed to discretize the angle direction of the microscopic transport equation. 
In addition, the integral solution of the transport equation is employed to establish the time-dependent interface fluxes at a cell interface for both micro and macroscopic equations. 
This solution covers the physics from the free transport to the diffusion limit, which makes the UGKS asymptotic preserving (AP). 
The UGKS accurately captures the diffusive and free transport solutions in optically thick and thin regime respectively, 
as well as the solutions in the transition regime. 
Recently, the UGKS is extended to the particle-based Monte Carlo method, 
such as the unified gas kinetic particle (UGKP) \cite{shi2020asymptotic,shi2021improved} and the unified gas-kinetic wave particle (UGKWP) methods \cite{li2020unified}. 
Similar to the UGKS framework, the UGKP and UGKWP methods employs a finite volume solver for the macroscopic transport and material energy equations and a particle-based Monte Carlo solver (instead of the DOM method) for the microscopic transport equation.

In this work, we extended the UGKP method for the frequency-dependent radiative system 
considering both absorption-emission and scattering process, 
and the scheme is formulated on a general unstructured mesh. 
The key methodology of the construction of UGKP is first the coupling evolution of macroscopic energy equation and microscopic radiant intensity equation, and second the multiscale numerical flux and closure derived from the integral solution.
The multi-frequency formulation is used to discretize the frequency space, 
and a multi-dimensional formulation is used for flux construction at cell interface. 
The photon particles are sampled and tracked from multiple sources, 
including photons in census, photons from boundary/initial condition and the macroscopic emission and scattering sources.
The proposed scheme is capable to capture the multiscale flow physics in both spatial and frequency space.
With the inclusion of scattering effect, the flow regimes are enriched, 
covering the optically thin ballistic regime, optically thick diffusive regime, and two-temperature diffusive regime \cite{godillon2005coupled,buet2004asymptotic}.
The proposed UGKP preserves all three regime solutions in their corresponding flow regimes.
Especially, the UGKP converges to a nine-point scheme \cite{sun2017multidimensional,sheng2008nine} 
on a distorted quadrilateral mesh in the diffusive regime. 

The rest of this paper is organized as follows. 
In section \ref{RTE}, we briefly introduce the TRT equations. 
In section \ref{UGKP}, the proposed UGKP method for solving the frequency-dependent radiative equations with considered both absorption-emission and scattering processes on unstructured mesh is presented. 
In section \ref{analysis}, the numerical properties including the asymptotic preserving property of UGKP are discussed.
Numerical results are shown in section \ref{examples} to verify the UGKP method and the program. 
The summary and future work are given in section \ref{conclusions}.

\section{Frequency-dependent radiative transfer equations}\label{RTE}
The frequency-dependent radiative transfer equations describe the transport of radiation and its energy exchange with material. 
Under the assumption of local thermal equilibrium (LTE), the time dependent frequency-dependent radiative transfer equations in the absence of both external and internal sources can be written in following scaled form:
\begin{equation} \label{Equation_2_1_}
\left\{\begin{array}{l} {\frac{\varepsilon }{c} \frac{\partial I}{\partial t} +\vec{\Omega }\cdot \nabla I=L_{a}^{\varepsilon } \sigma _{a} \left(B\left(\nu ,T\right)-I\right)+L_{s}^{\varepsilon } \sigma _{s} \left(\frac{1}{4\pi } \int _{4\pi }Id\vec{\Omega } -I\right)} \\ {C_{V} \frac{\partial T}{\partial t} \equiv \frac{\partial U_{m} }{\partial t} =\frac{L_{a}^{\varepsilon } \sigma _{a} }{\varepsilon } \int _{4\pi }\int _{0}^{\infty }\left(I-B\left(\nu ,T\right)\right)d\nu d\vec{\Omega }  } \end{array}\right.
\end{equation}
Here $I\left(t,\vec{r},\vec{\Omega },\nu \right)$ is the radiation intensity which depends on time $t$, 
spatial variable $\vec{r}$, 
angular variable $\vec{\Omega }$, 
frequency variable $\nu \in \left(0,+\infty \right)$, $T\left(t,\vec{r}\right)$ is the material temperature, 
$\sigma _{a} \left(\vec{r},\nu ,T\right)$ is the absorption coefficient, 
$\sigma _{s} \left(\vec{r},\nu ,T\right)$ is the scattering coefficient, 
$c$ is the speed of light, $\varepsilon >0$ is the Knudsen number, 
$L_{a}^{\varepsilon } $ and $L_{s}^{\varepsilon } $ are two parameters depending on $\varepsilon $, 
$U_{m} \left(\vec{r},t\right)$ is the material energy density, 
and $C_{V} >0$ is the heat capacity. 
In addition, the Planck function $B\left(\nu ,T\right)$ is defined by
\begin{equation} \label{Equation_2_2_}
B\left(\nu ,T\right)=\frac{2h\nu ^{3} }{c^{2} } \frac{1}{e^{{h\nu \mathord{\left/ {\vphantom {h\nu  kT}} \right. \kern-\nulldelimiterspace} kT} } -1}
\end{equation}
where $h$ is Planck's constant and $k$ is Boltzmann's constant. 
For simplicity, the assumption of isotropic scattering is taken in this work.

Since both scattering and absorption process are considered, 
system \eqref{Equation_2_1_} will relax to different equilibrium states in the diffusive regime depends on which process is dominated. 
With $L_{a}^{\varepsilon } ={1\mathord{\left/ {\vphantom {1 \varepsilon }} \right. \kern-\nulldelimiterspace} \varepsilon } $ and $L_{s}^{\varepsilon } =\varepsilon $, 
the absorption process will be dominated as $\varepsilon \to 0$, 
thus the equilibrium state with equal radiation and material temperature and Planckian distribution $I\to B\left(\nu ,T\right)$ will be obtained, 
and the material temperature $T_{0} $ satisfies the following evolution equation:
\begin{equation} \label{Equation_2_3_}
\frac{\partial }{\partial t} U_{m} \left(T_{0} \right)+\frac{\partial }{\partial t} \left(aT_{0}^{4} \right)=\nabla \cdot \frac{1}{3\sigma _{a} } \nabla \left(acT_{0}^{4} \right)
\end{equation}
where $a$ is the radiation constant given by
\[a=\frac{8\pi k^{4} }{15h^{3} c^{3} } \]
With $L_{a}^{\varepsilon } =\varepsilon $ and $L_{s}^{\varepsilon } ={1\mathord{\left/ {\vphantom {1 \varepsilon }} \right. \kern-\nulldelimiterspace} \varepsilon } $, 
however, the scattering process will be dominated as $\varepsilon \to 0$, and the non-equilibrium states will be approached. 
If we define radiation energy as $\rho =\int _{{\rm 4}\pi }Id\vec{\Omega } $, 
the radiation intensity goes to $I\to \frac{\rho }{4\pi } $, 
and the radiation energy and material temperature $T$ stratify the following nonlinear non-equilibrium diffusion equations:
\begin{equation} \label{Equation_2_4_}
\left\{\begin{array}{l} {\frac{\partial \rho }{\partial t} -\nabla \cdot \frac{c}{3\sigma _{s} } \nabla \rho =c\sigma _{a} \left(4\pi B\left(\nu ,T\right)-\rho \right)} \\ {C_{V} \frac{\partial T}{\partial t} \equiv \frac{\partial U_{m} }{\partial t} =\sigma _{a} \int _{0}^{\infty }\left(\rho -4\pi B\left(\nu ,T\right)\right)d\nu  } \end{array}\right.
\end{equation}

An asymptotic preserving (AP) scheme for the frequency-dependent radiation transfer system \eqref{Equation_2_1_} is a numerical scheme which should lead to the correct discretization of different equilibrium states under different condition at small $\varepsilon $ case. 
At the same time, the AP scheme should be uniformly stable in $\varepsilon $.

\section{ UGKP for frequency-dependent radiative transfer system}\label{UGKP}
In this section, we present the UGKP method for the frequency-dependent radiative equations with both absorption-emission and scattering process under unstructured mesh.
\subsection{Frequency space discretization}
First, we discretize the frequency variable $\nu $ for the system \eqref{Equation_2_1_} with the standard multi-group method. 
In the multi-group method, the frequency variable $\nu $ is divided into discrete frequency intervals and groups the photons according to these intervals. 
Here we discretize the continuous frequency space $\left({\rm 0,+}\infty \right)$ into $G$ discrete frequency internals with frequency boundary $\left(\nu _{g-{1\mathord{\left/ {\vphantom {1 2}} \right. \kern-\nulldelimiterspace} 2} } {\rm ,}\nu _{g+{1\mathord{\left/ {\vphantom {1 2}} \right. \kern-\nulldelimiterspace} 2} } \right)$, where $g=1,\ldots ,G$, and $\nu _{{1\mathord{\left/ {\vphantom {1 2}} \right. \kern-\nulldelimiterspace} 2} } =0,\nu _{{G+1\mathord{\left/ {\vphantom {G+1 2}} \right. \kern-\nulldelimiterspace} 2} } =\infty $. 
With defined groups boundaries above, we can integrate the first equation in \eqref{Equation_2_1_} over each frequency interval:
\begin{equation} \label{Equation_3_1_}
\int _{\nu _{g-{1\mathord{\left/ {\vphantom {1 2}} \right. \kern-\nulldelimiterspace} 2} } }^{\nu _{g+{1\mathord{\left/ {\vphantom {1 2}} \right. \kern-\nulldelimiterspace} 2} } }\left(\frac{\varepsilon }{c} \frac{\partial I}{\partial t} +\vec{\Omega }\cdot \nabla I\right)d\nu  =\int _{\nu _{g-{1\mathord{\left/ {\vphantom {1 2}} \right. \kern-\nulldelimiterspace} 2} } }^{\nu _{g+{1\mathord{\left/ {\vphantom {1 2}} \right. \kern-\nulldelimiterspace} 2} } }\left[L_{a}^{\varepsilon } \sigma _{a} \left(B\left(\nu ,T\right)-I\right)\right]d\nu  +\int _{\nu _{g-{1\mathord{\left/ {\vphantom {1 2}} \right. \kern-\nulldelimiterspace} 2} } }^{\nu _{g+{1\mathord{\left/ {\vphantom {1 2}} \right. \kern-\nulldelimiterspace} 2} } }\left[L_{s}^{\varepsilon } \sigma _{s} \left(\frac{1}{4\pi } \int _{4\pi }Id\vec{\Omega } -I\right)\right]d\nu
\end{equation}
For equation \eqref{Equation_3_1_}, the radiation intensity and energy in different groups and the corresponding group opacities are given by
\begin{equation} \label{Equation_3_2_}
I_{g} =\int _{\nu _{g-{1\mathord{\left/ {\vphantom {1 2}} \right. \kern-\nulldelimiterspace} 2} } }^{\nu _{g+{1\mathord{\left/ {\vphantom {1 2}} \right. \kern-\nulldelimiterspace} 2} } }I\left(t,\vec{r},\vec{\Omega },\nu \right)d\nu  ,\; \; \rho _{g} =\int _{{\rm 4}\pi }I_{g} d\vec{\Omega }
\end{equation}
and
\begin{equation} \label{Equation_3_3_}
\begin{array}{l} {\sigma _{e,g} =\frac{\int _{\nu _{g-{1\mathord{\left/ {\vphantom {1 2}} \right. \kern-\nulldelimiterspace} 2} } }^{\nu _{g+{1\mathord{\left/ {\vphantom {1 2}} \right. \kern-\nulldelimiterspace} 2} } }\sigma _{a} B\left(\nu ,T\right)d\nu  }{\int _{\nu _{g-{1\mathord{\left/ {\vphantom {1 2}} \right. \kern-\nulldelimiterspace} 2} } }^{\nu _{g+{1\mathord{\left/ {\vphantom {1 2}} \right. \kern-\nulldelimiterspace} 2} } }B\left(\nu ,T\right)d\nu  } ,\sigma _{a,g} =\frac{\int _{\nu _{g-{1\mathord{\left/ {\vphantom {1 2}} \right. \kern-\nulldelimiterspace} 2} } }^{\nu _{g+{1\mathord{\left/ {\vphantom {1 2}} \right. \kern-\nulldelimiterspace} 2} } }\sigma _{a} Id\nu  }{\int _{\nu _{g-{1\mathord{\left/ {\vphantom {1 2}} \right. \kern-\nulldelimiterspace} 2} } }^{\nu _{g+{1\mathord{\left/ {\vphantom {1 2}} \right. \kern-\nulldelimiterspace} 2} } }Id\nu  } ,} \\ {\sigma _{s-out,g} =\frac{\int _{\nu _{g-{1\mathord{\left/ {\vphantom {1 2}} \right. \kern-\nulldelimiterspace} 2} } }^{\nu _{g+{1\mathord{\left/ {\vphantom {1 2}} \right. \kern-\nulldelimiterspace} 2} } }\sigma _{s} Id\nu  }{\int _{\nu _{g-{1\mathord{\left/ {\vphantom {1 2}} \right. \kern-\nulldelimiterspace} 2} } }^{\nu _{g+{1\mathord{\left/ {\vphantom {1 2}} \right. \kern-\nulldelimiterspace} 2} } }Id\nu  } ,\sigma _{s-in,g} =\frac{\int _{\nu _{g-{1\mathord{\left/ {\vphantom {1 2}} \right. \kern-\nulldelimiterspace} 2} } }^{\nu _{g+{1\mathord{\left/ {\vphantom {1 2}} \right. \kern-\nulldelimiterspace} 2} } }\sigma _{s} \left(\int _{4\pi }Id\vec{\Omega } \right)d\nu  }{\int _{\nu _{g-{1\mathord{\left/ {\vphantom {1 2}} \right. \kern-\nulldelimiterspace} 2} } }^{\nu _{g+{1\mathord{\left/ {\vphantom {1 2}} \right. \kern-\nulldelimiterspace} 2} } }\left(\int _{4\pi }Id\vec{\Omega } \right)d\nu  } } \end{array}
\end{equation}
For Planck function $B\left(\nu ,T\right)$ on the right side of equation \eqref{Equation_3_1_}, it is also integrated over the frequency interval by
\begin{equation} \label{Equation_3_4_}
\phi _{g} =\int _{\nu _{g-{1\mathord{\left/ {\vphantom {1 2}} \right. \kern-\nulldelimiterspace} 2} } }^{\nu _{g+{1\mathord{\left/ {\vphantom {1 2}} \right. \kern-\nulldelimiterspace} 2} } }B\left(\nu ,T\right)d\nu
\end{equation}
With these notations in \eqref{Equation_3_2_}, \eqref{Equation_3_3_} and \eqref{Equation_3_4_}, equation \eqref{Equation_2_1_} turns to an equivalent multi-group radiative transfer system.
\begin{equation} \label{Equation_3_5_}
\left\{\begin{array}{l} {\frac{\varepsilon }{c} \frac{\partial I_{g} }{\partial t} +\vec{\Omega }\cdot \nabla I_{g} =L_{a}^{\varepsilon } \left(\sigma _{e,g} \phi _{g} -\sigma _{a,g} I_{g} \right)+L_{s}^{\varepsilon } \left(\sigma _{s-in,g} \frac{\rho _{g} }{4\pi } -\sigma _{s-out,g} I_{g} \right)} \\ {C_{V} \frac{\partial T}{\partial t} \equiv \frac{\partial U_{m} }{\partial t} =\frac{L_{a}^{\varepsilon } }{\varepsilon } \sum _{g=1}^{G}\int _{4\pi }\left(\sigma _{a,g} I_{g} -\sigma _{e,g} \phi _{g} \right)d\vec{\Omega }  } \end{array}\right.
\end{equation}
One should note that the absorption opacity $\sigma _{a,g} $, in-scattering opacity $\sigma _{s-in,g} $ and out-scattering opacity $\sigma _{s-out,g} $ in equation \eqref{Equation_3_5_} is a weighted integration with the unknown function \textit{I}. Usually, the unknown function \textit{I} in these opacity integration is replaced by the Planck function with radiation temperature $T_{r} $ in the following ways:
\begin{equation} \label{Equation_3_6_}
\begin{array}{l} {acT_{r}^{4} =\int _{4\pi }\int _{0}^{\infty }Id\nu d\vec{\Omega } =\sum _{g=1}^{G}\int _{4\pi }I_{g} d\vec{\Omega }   =\sum _{g=1}^{G}\rho _{g}  ,} \\ {\sigma _{a,g} =\frac{\int _{\nu _{g-{1\mathord{\left/ {\vphantom {1 2}} \right. \kern-\nulldelimiterspace} 2} } }^{\nu _{g+{1\mathord{\left/ {\vphantom {1 2}} \right. \kern-\nulldelimiterspace} 2} } }\sigma _{a} B\left(\nu ,T_{r} \right)d\nu  }{\int _{\nu _{g-{1\mathord{\left/ {\vphantom {1 2}} \right. \kern-\nulldelimiterspace} 2} } }^{\nu _{g+{1\mathord{\left/ {\vphantom {1 2}} \right. \kern-\nulldelimiterspace} 2} } }B\left(\nu ,T_{r} \right)d\nu  } ,\; \; \sigma _{s-out,g} =\frac{\int _{\nu _{g-{1\mathord{\left/ {\vphantom {1 2}} \right. \kern-\nulldelimiterspace} 2} } }^{\nu _{g+{1\mathord{\left/ {\vphantom {1 2}} \right. \kern-\nulldelimiterspace} 2} } }\sigma _{s} B\left(\nu ,T_{r} \right)d\nu  }{\int _{\nu _{g-{1\mathord{\left/ {\vphantom {1 2}} \right. \kern-\nulldelimiterspace} 2} } }^{\nu _{g+{1\mathord{\left/ {\vphantom {1 2}} \right. \kern-\nulldelimiterspace} 2} } }B\left(\nu ,T_{r} \right)d\nu  } ,} \\ {\sigma _{s-in,g} =\frac{\int _{\nu _{g-{1\mathord{\left/ {\vphantom {1 2}} \right. \kern-\nulldelimiterspace} 2} } }^{\nu _{g+{1\mathord{\left/ {\vphantom {1 2}} \right. \kern-\nulldelimiterspace} 2} } }\sigma _{s} \left(\int _{4\pi }B\left(\nu ,T_{r} \right)d\vec{\Omega } \right)d\nu  }{\int _{\nu _{g-{1\mathord{\left/ {\vphantom {1 2}} \right. \kern-\nulldelimiterspace} 2} } }^{\nu _{g+{1\mathord{\left/ {\vphantom {1 2}} \right. \kern-\nulldelimiterspace} 2} } }\left(\int _{4\pi }B\left(\nu ,T_{r} \right)d\vec{\Omega } \right)d\nu  } =\frac{4\pi \int _{\nu _{g-{1\mathord{\left/ {\vphantom {1 2}} \right. \kern-\nulldelimiterspace} 2} } }^{\nu _{g+{1\mathord{\left/ {\vphantom {1 2}} \right. \kern-\nulldelimiterspace} 2} } }\sigma _{s} B\left(\nu ,T_{r} \right)d\nu  }{4\pi \int _{\nu _{g-{1\mathord{\left/ {\vphantom {1 2}} \right. \kern-\nulldelimiterspace} 2} } }^{\nu _{g+{1\mathord{\left/ {\vphantom {1 2}} \right. \kern-\nulldelimiterspace} 2} } }B\left(\nu ,T_{r} \right)d\nu  } =\sigma _{s-out,g} } \end{array}
\end{equation}
Thus, we set $\sigma _{s,g} =\sigma _{s-in,g} =\sigma _{s-out,g} $ as the scattering opacity, and the equation \eqref{Equation_3_5_} turns to:
\begin{equation} \label{Equation_3_7_}
\left\{\begin{array}{l} {\frac{\varepsilon }{c} \frac{\partial I_{g} }{\partial t} +\vec{\Omega }\cdot \nabla I_{g} =L_{a}^{\varepsilon } \left(\sigma _{e,g} \phi _{g} -\sigma _{a,g} I_{g} \right)+L_{s}^{\varepsilon } \sigma _{s,g} \left(\frac{\rho _{g} }{4\pi } -I_{g} \right)} \\ {C_{V} \frac{\partial T}{\partial t} \equiv \frac{\partial U_{m} }{\partial t} =\frac{L_{a}^{\varepsilon } }{\varepsilon } \sum _{g=1}^{G}\int _{4\pi }\left(\sigma _{a,g} I_{g} -\sigma _{e,g} \phi _{g} \right)d\vec{\Omega }  } \end{array}\right.
\end{equation}
Taking the angular integration of the radiation transport equation in \eqref{Equation_3_7_}, the following macroscopic equations can be obtained.
\begin{equation} \label{Equation_3_8_}
\left\{\begin{array}{l} {\frac{\varepsilon }{c} \frac{\partial \rho _{g} }{\partial t} +\nabla \cdot \left\langle \vec{\Omega }I_{g} \right\rangle =L_{a}^{\varepsilon } \left(4\pi \sigma _{e,g} \phi _{g} -\sigma _{a,g} \rho _{g} \right)} \\ {C_{V} \frac{\partial T}{\partial t} \equiv \frac{\partial U_{m} }{\partial t} =\frac{L_{a}^{\varepsilon } }{\varepsilon } \sum _{g=1}^{G}\left(\sigma _{a,g} \rho _{g} -4\pi \sigma _{e,g} \phi _{g} \right) } \end{array}\right.
\end{equation}
where $\left\langle \vec{\Omega }I_{g} \right\rangle {\rm =}\int _{{\rm 4}\pi }\vec{\Omega }I_{g} d\vec{\Omega } $. Up to now, the discretization of the frequency variable is finished.

\subsection{Spatial and time discretization}
Under the multi-group framework, we will give the discretization of spatial and time variable on unstructured mesh here, based on the finite volume method. For simplicity, we only consider the two-dimensional Cartesian spatial case for problem \eqref{Equation_2_1_}. But the extension to three-dimensional case is straightforward.

 In this case, the angle direction is denoted by $\vec{\Omega }{\rm =}\left(\mu ,\xi \right)$ with $\mu =\sqrt{1-\zeta ^{2} } \cos \theta $ and$\xi =\sqrt{1-\zeta ^{2} } \sin \theta $, where $\zeta \in \left[-1,1\right]$ is the cosine value of the angle between the propagation direction $\vec{\Omega }$ and the \textit{z}-axis, and $\theta \in \left[0,2\pi \right)$ is the angle between the projection vector of $\vec{\Omega }$ onto the \textit{xy}-plane and the \textit{x}-axis. Due to the symmetry of angular distribution in the two-dimensional Cartesian case, we only need to consider $\zeta >0$.

For the spatial variables, the unstructured mesh in two dimensional case is used, where the computational mesh is assumed to be composed of quadrilaterals. Figure 1 shows a generalize 2D quadrilateral cell \textit{j }whose area will be denoted by $V_{j} $. Its center $c_{j} =\left(x_{j}^{c} ,y_{j}^{c} \right)$ is given by
\begin{equation} \label{Equation_3_9_}
\left\{\begin{array}{l} {x_{j}^{c} =\frac{1}{V_{j} } \int _{V_{j} }xdxdy } \\ {y_{j}^{c} =\frac{1}{V_{j} } \int _{V_{j} }ydxdy } \end{array}\right.
\end{equation}
And the time is discretized by $t_{n} $ with time step $\Delta t=t_{n+1} -t_{n} $. Thus, a conservative finite volume numerical scheme for the macroscopic equation \eqref{Equation_3_8_} is of the form
\begin{equation} \label{Equation_3_10_}
\left\{\begin{array}{l} {\rho _{j,g}^{n{\rm +1}} =\rho _{j,g}^{n} -\frac{\Delta t}{V_{j} } \sum _{k}\Phi _{j,k,g}^{n+1}  +\frac{c\Delta tL_{a}^{\varepsilon } }{\varepsilon } \left(2\pi \left(\sigma _{e,g} \right)_{j}^{n+1} \phi _{j,g}^{n{\rm +1}} -\left(\sigma _{a,g} \right)_{j}^{n+1} \rho _{j,g}^{n{\rm +1}} \right)} \\ {C_{V} \frac{T_{j}^{n+1} -T_{j}^{n} }{\Delta t} =\frac{L_{a}^{\varepsilon } }{\varepsilon } \sum _{g=1}^{G}\left(\left(\sigma _{a,g} \right)_{j}^{n+1} \rho _{j,g}^{n{\rm +1}} -2\pi \left(\sigma _{e,g} \right)_{j}^{n+1} \phi _{j,g}^{n{\rm +1}} \right) } \end{array}\right.
\end{equation}
where $\rho _{j,g}^{n+1} $, $\phi _{j,g}^{n+1} $ and $T_{j}^{n+1} $ are the cell averaged value at time $t_{n} $ in cell $j$, and $\Phi _{j,k,g}^{n+1} $ is the macroscopic fluxes across the cell edge \textit{k}. The explicit expressions for these terms are given in the following:
\begin{equation} \label{Equation_3_11_}
\begin{array}{l} {\; \rho _{j,g}^{n+1} =\frac{1}{V_{j} } \int _{V_{j} }\rho _{g} \left(t_{n+1} ,x,y\right)dxdy ,{\rm \; \; }\phi _{j,g}^{n+1} =\frac{1}{V_{j} } \int _{V_{j} }\phi _{g} \left(t_{n+1} ,x,y\right)dxdy ,} \\ {\Phi _{j,k,g}^{n+1} =\frac{cl_{k} }{\varepsilon \Delta t} \int _{t_{n} }^{t_{n+1} }\vec{n}_{k} \cdot \left\langle \vec{\Omega }I_{g} \right\rangle \left(t,p_{k,m} \right)dt } \end{array}
\end{equation}
where $p_{k,m} $ is the center of edge $k$, and $\vec{n}_{k} $ is the unit normal vector of edge \textit{k}. The construction of $\Phi _{j,k,g}^{n+1} $ is the key of UGKP method, which will be discussed in the next subsection.

\begin{figure}
  \centering
  \includegraphics[width=0.6\textwidth]{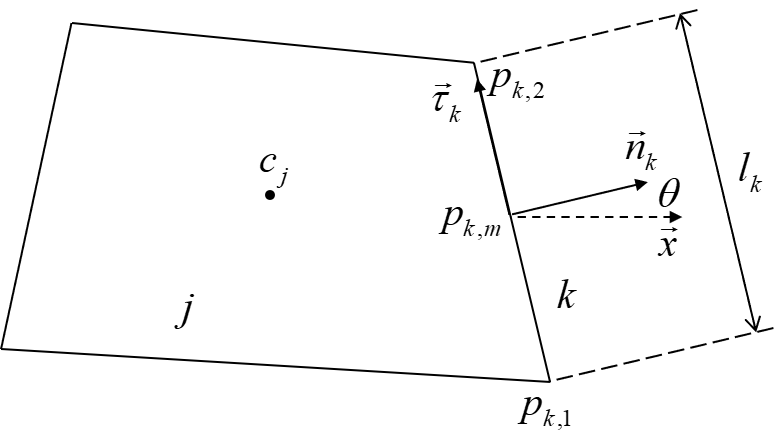}
  \caption{A cell $j$ of the generalize quadrilateral mesh with $c_{j} $ as the cell center, $p_{k,m} $ as the center of edge $k$. The length of the edge $k$ is $l_{k} $. The tow vertexes of edge $k$ are $\vec{p}_{k,1} $ and $\vec{p}_{k,2} $, and the unit normal and tangential vector is $\vec{n}_{k} $ and $\vec{\tau }_{k} $ respectively. $\theta $ is the angle between $\vec{n}_{k} $ and the \textit{x}-axis.}
  \label{fig_image1}
\end{figure}

\subsection{Construction of the macroscopic fluxes $\Phi _{j,k,g}^{n+1} $}
With the methodology of UGKS, the radiation intensity $I_{g} $ around the center of edge \textit{k} can be obtained by solving the following initial value problem:
\begin{equation} \label{Equation_3_12_}
\left\{\begin{array}{l} {\frac{\varepsilon }{c} \frac{\partial I_{g} }{\partial t} +\mu \frac{\partial I_{g} }{\partial x} +\xi \frac{\partial I_{g} }{\partial y} =L_{a}^{\varepsilon } \left(\sigma _{e,g} \phi _{g} -\sigma _{a,g} I_{g} \right)+L_{s}^{\varepsilon } \sigma _{s,g} \left(\frac{\rho _{g} }{2\pi } -I_{g} \right)} \\ {\left. I_{g} \left(x,y,t\right)\right|_{t=t_{n} } =I_{g} \left(x,y,t_{n} \right)} \end{array}\right.
\end{equation}

For an unstructured mesh, we need consider the flow variation in both normal and tangential direction ($\vec{n}_{k} $ and $\vec{\tau }_{k} $) of edge \textit{k}. The angle between the normal direction $\vec{n}_{k} $ and the positive global \textit{x}-axis $\vec{x}$ is $\theta $, as shown in Figure 1. Thus, we have the following relationships:
\begin{equation} \label{Equation_3_13_}
\left\{\begin{array}{l} {\vec{n}_{k} =\left(\cos \theta ,\sin \theta \right)} \\ {\vec{\tau }_{k} =\left(-\sin \theta ,\cos \theta \right)} \end{array}\right.
\end{equation}
With equations \eqref{Equation_3_13_}, we could transfer the global coordinate $\left(x,y\right)$ to local orthogonal coordinate $\left(x',y'\right)$ by
\begin{equation} \label{Equation_3_14_}
\left\{\begin{array}{l} {x'=\left(x-x_{j}^{k,m} \right)\cos \theta +\left(y-y_{j}^{k,m} \right)\sin \theta } \\ {y'=-\left(x-x_{j}^{k,m} \right)\sin \theta +\left(y-y_{j}^{k,m} \right)\cos \theta } \end{array}\right.
\end{equation}
where $x_{j}^{k,m} $ and $y_{j}^{k,m} $ is the global coordinate of edge center $p_{k,m} $. Then the equation \eqref{Equation_3_12_} could be written as
\begin{equation} \label{Equation_3_15_}
\left\{\begin{array}{l} {\frac{\varepsilon }{c} \frac{\partial I_{g} }{\partial t} +\mu '\frac{\partial I_{g} }{\partial x'} +\xi '\frac{\partial I_{g} }{\partial y'} =L_{a}^{\varepsilon } \left(\sigma _{e,g} \phi _{g} -\sigma _{a,g} I_{g} \right)+L_{s}^{\varepsilon } \sigma _{s,g} \left(\frac{\rho _{g} }{2\pi } -I_{g} \right)} \\ {\left. I_{g} \left(x',y',t\right)\right|_{t=t_{n} } =I_{g} \left(x',y',t_{n} \right)} \end{array}\right.
\end{equation}
where $\left(\mu ',\xi '\right)$ is transformed from $\left(\mu ,\xi \right)$ with a similar relation as \eqref{Equation_3_14_}.

With equation \eqref{Equation_3_14_} and \eqref{Equation_3_15_}, the following integral solution can be obtained around the edge center $p_{k,m} $:
\begin{equation} \label{Equation_3_16_}
\begin{array}{l} {I_{g} \left(t,{\rm 0},{\rm 0}\right)=e^{-\lambda _{g} \left(t-t_{n} \right)} I_{g} \left(t_{n} ,-\frac{\mu 'c}{\varepsilon } \left(t-t_{n} \right),-\frac{\xi 'c}{\varepsilon } \left(t-t_{n} \right)\right)} \\ {+\int _{t_{n} }^{t}e^{-\lambda _{g} \left(t-s\right)} \left[\begin{array}{l} {\frac{cL_{a}^{\varepsilon } \sigma _{e,g} }{\varepsilon } \phi _{g} \left(s,-\frac{\mu 'c}{\varepsilon } \left(t-s\right),-\frac{\xi 'c}{\varepsilon } \left(t-s\right)\right)} \\ {+\frac{cL_{s}^{\varepsilon } \sigma _{s,g} }{\varepsilon } \frac{\rho _{g} }{{\rm 2}\pi } \left(s,-\frac{\mu 'c}{\varepsilon } \left(t-s\right),-\frac{\xi 'c}{\varepsilon } \left(t-s\right)\right)} \end{array}\right]ds } \end{array}
\end{equation}
where $\lambda _{g} =\frac{c\left(L_{a}^{\varepsilon } \sigma _{a,g} +L_{s}^{\varepsilon } \sigma _{s,g} \right)}{\varepsilon } $. And the macroscopic flux across the edge \textit{k} can be evaluated by
\begin{equation} \label{Equation_3_17_}
\Phi _{j,k,g}^{n+1} =\frac{cl_{k} }{\varepsilon \Delta t} \int _{t_{n} }^{t_{n+1} }\left\langle \mu 'I_{g} \right\rangle \left(t,0,0\right)dt
\end{equation}

The first part of equation \eqref{Equation_3_16_} contributes from microscopic photon free transport, including both the initial/boundary photons and the free photons from previous time step in the computational domain. While the second part of equation \eqref{Equation_3_16_} denotes the contribution of macroscopic emission and scattering photons in this time step. Thus, the microscopic and macroscopic terms could play different roles in different transport regimes in this integral solution \eqref{Equation_3_16_}, making \eqref{Equation_3_17_} as a multiscale numerical flux in the UGKP method.

However, the function $\phi _{g} $ and $\rho _{g} $ are still needed to get the explicit expressions of \eqref{Equation_3_16_} and \eqref{Equation_3_17_}. We model it by a piecewise continuous polynomial around the edge \textit{k} as follows:
\begin{equation} \label{Equation_3_18_}
\begin{array}{l} {\phi _{g} \left(t,x',y'\right)=\phi _{j,k,g}^{n+1} +\delta _{t} \phi _{j,k,g}^{n+1} \left(t-t_{n+1} \right)+\delta _{y'} \phi _{j,k,g}^{n+1} y'+\delta _{x'} \phi _{j,k,g}^{n+1} x'} \\ {\rho _{g} \left(t,x',y'\right)=\rho _{j,k,g}^{n+1} +\delta _{t} \rho _{j,k,g}^{n+1} \left(t-t_{n+1} \right)+\delta _{y'} \rho _{j,k,g}^{n+1} y'+\delta _{x'} \rho _{j,k,g}^{n+1} x'} \end{array}
\end{equation}
The spatial derivatives in \eqref{Equation_3_18_} has following formation:
\begin{equation} \label{Equation_3_19_}
\begin{array}{l} {\delta _{x'} \phi _{j,k,g}^{n+1} =\frac{\phi _{j',g}^{n+1} -\phi _{j,g}^{n+1} -\left(\tau _{j,k}^{-} +\tau _{j,k}^{+} \right)\left(\hat{\phi }_{k,2,g}^{n+1} -\hat{\phi }_{k,1,g}^{n+1} \right)}{l_{j,k}^{-} +l_{j,k}^{+} } ,\; \; \delta _{y'} \phi _{j,k,g}^{n+1} =\frac{\hat{\phi }_{k,2,g}^{n+1} -\hat{\phi }_{k,1,g}^{n+1} }{l_{k} } } \\ {\delta _{x'} \rho _{j,k,g}^{n+1} =\frac{\rho _{j',g}^{n+1} -\rho _{j,g}^{n+1} -\left(\tau _{j,k}^{-} +\tau _{j,k}^{+} \right)\left(\hat{\rho }_{k,2,g}^{n+1} -\hat{\rho }_{k,1,g}^{n+1} \right)}{l_{j,k}^{-} +l_{j,k}^{+} } ,\; \; \delta _{y'} \rho _{j,k,g}^{n+1} =\frac{\hat{\rho }_{k,2,g}^{n+1} -\hat{\rho }_{k,1,g}^{n+1} }{l_{k} } } \end{array}
\end{equation}
where $j'$ denotes the neighboring cell which has the common edge \textit{k} with cell \textit{j}. The $\hat{\phi }_{k,1,g}^{n+1} ,\; \; \hat{\phi }_{k,{\rm 2,g}}^{n+1} $and$\hat{\rho }_{k,1,g}^{n+1} ,\; \; \hat{\rho }_{k,2,g}^{n+1} $ values are the macroscopic quantities on the two vertexes ($p_{k,1} $ and $p_{k,2} $) of edge $k$, which is calculated as the average macroscopic quantities of those cells have the common vertex. And other terms on the right side of equation \eqref{Equation_3_19_} are the projected length given by
\begin{equation} \label{Equation_3_20_}
\left\{\begin{array}{l} {l_{j,k}^{-} =\vec{r}\left(c_{j} ,p_{k,m} \right)\cdot \vec{n}_{k} } \\ {l_{j,k}^{+} =\vec{r}\left(p_{k,m} ,c_{j'} \right)\cdot \vec{n}_{k} } \\ {\tau _{j,k}^{-} ={\left[\vec{r}\left(c_{j} ,p_{k,m} \right)\cdot \vec{\tau }_{k} \right]\mathord{\left/ {\vphantom {\left[\vec{r}\left(c_{j} ,p_{k,m} \right)\cdot \vec{\tau }_{k} \right] l_{k} }} \right. \kern-\nulldelimiterspace} l_{k} } } \\ {\tau _{j,k}^{+} ={\left[\vec{r}\left(p_{k,m} ,c_{j'} \right)\cdot \vec{\tau }_{k} \right]\mathord{\left/ {\vphantom {\left[\vec{r}\left(p_{k,m} ,c_{j'} \right)\cdot \vec{\tau }_{k} \right] l_{k} }} \right. \kern-\nulldelimiterspace} l_{k} } } \end{array}\right.
\end{equation}
In addition, the time derivative in \eqref{Equation_3_18_} is given by
\begin{equation} \label{Equation_3_21_}
\delta _{t} \phi _{j,k,g}^{n+1} =\frac{\phi _{j,k,g}^{n+1} -\phi _{j,k,g}^{n} }{\Delta t} ,\; \; \delta _{t} \rho _{j,k,g}^{n+1} =\frac{\rho _{j,k,g}^{n+1} -\rho _{j,k,g}^{n} }{\Delta t}
\end{equation}
Based on the above notations, the macroscopic interface flux $\Phi _{j,k,g}^{n+1} $ can be exactly computed by substituting \eqref{Equation_3_16_} into \eqref{Equation_3_17_}, which has the following formulation:
\begin{equation} \label{Equation_3_22_}
\begin{array}{l} {\Phi _{j,k,g}^{n+1} =\left\langle \frac{cl_{k} }{\varepsilon \Delta t} \int _{t_{n} }^{t_{n+1} }\mu 'e^{-\lambda _{g} \left(t-t_{n} \right)} I_{g} \left(t_{n} ,-\frac{\mu 'c}{\varepsilon } \left(t-t_{n} \right),-\frac{\xi 'c}{\varepsilon } \left(t-t_{n} \right)\right)dt \right\rangle } \\ {+\frac{2\pi }{3} \left(D^{1} \right)_{j,k,g}^{n+1} \delta _{x'} \phi _{j,k,g}^{n+1} +\frac{2\pi }{3} \left(D^{2} \right)_{j,k,g}^{n+1} \delta _{x'} \rho _{j,k,g}^{n+1} } \end{array}
\end{equation}
The coefficients in \eqref{Equation_3_22_} are given by
\begin{equation} \label{Equation_3_23_}
D^{1} =-\frac{c^{{\rm 3}} l_{k} L_{a}^{\varepsilon } \sigma _{e,g} }{\Delta t\varepsilon ^{{\rm 3}} \lambda _{g}^{2} } \left[\Delta t\left(1+e^{-\lambda _{g} \Delta t} \right)-\frac{2}{\lambda _{g} } \left(1-e^{-\lambda _{g} \Delta t} \right)\right],{\rm \; \; \; }D^{2} =-\frac{c^{{\rm 3}} l_{k} L_{s}^{\varepsilon } \sigma _{s,g} }{2\pi \Delta t\varepsilon ^{{\rm 3}} \lambda _{g}^{2} } \left[\Delta t\left(1+e^{-\lambda _{g} \Delta t} \right)-\frac{2}{\lambda _{g} } \left(1-e^{-\lambda _{g} \Delta t} \right)\right]
\end{equation}
The expressions \eqref{Equation_3_23_} have functional dependence on physical coefficients, the time step, and asymptotic parameter:
\begin{equation} \label{Equation_3_24_}
\left(D^{1} \right)_{j,k,g}^{n+1} =D^{1} \left(\Delta t,\varepsilon ,\left(\sigma _{e,g} \right)_{j,k,g}^{n+1} ,\left(\lambda _{g} \right)_{j,k,g}^{n+1} \right),\; \; \left(D^{2} \right)_{j,k,g}^{n+1} =D^{2} \left(\Delta t,\varepsilon ,\left(\sigma _{s,g} \right)_{j,k,g}^{n+1} ,\left(\lambda _{g} \right)_{j,k,g}^{n+1} \right)
\end{equation}
where the coefficients at cell edge are defined using the values from neighboring cells. Similar to \eqref{Equation_3_16_}, equation \eqref{Equation_3_22_} has also two contributions from both microscopic term (the angular integral term) and macroscopic terms (the last two terms). These macroscopic quantities will be discussed in the subsection 3.4, while the microscopic term will be obtained by the particle-based Monte Carlo solver in the subsection 3.5.

\subsection{Macroscopic solver for the radiation energy and material temperature}
In this subsection, we will show how to determine the macroscopic terms in \eqref{Equation_3_16_} and \eqref{Equation_3_22_} by solving the macroscopic equations \eqref{Equation_3_10_}, following the methodology of UGKP.

Substituting \eqref{Equation_3_22_} into the macroscopic equations \eqref{Equation_3_10_}, we get a coupled nonlinear system for the macroscopic quantities $T_{j}^{n+1} $ and $\rho _{j,g}^{n+1} $, with other quantities and parameters depended implicitly on $T_{j}^{n+1} $ and $\rho _{j,g}^{n+1} $. However, the microscopic term in \eqref{Equation_3_22_} which contributed from free photon transport is an unknown quantity here. We assume that the microscopic quantity is already known for now, and we will discuss about this in the next subsection. Thus, this nonlinear system can be solved by an iterative method, which is given below.

\textbf{Algorithm for solving (3.10).}
\begin{enumerate}
\item Based on the radiation energy $\rho _{j,g}^{n} $ and material temperature $T_{j}^{n} $ from the previous time step, we have the macroscopic quantity $\phi _{j,g}^{n} $. Then, we set the initial iterative value $\rho _{j,g}^{n+1,0} =\rho _{j,g}^{n} $ and $T_{j}^{n+1,0} =T_{j}^{n} $;
\item  For iteration $s=0,\ldots ,S$
\end{enumerate}

2.1) Compute the coefficients $\left(\sigma _{e,g} \right)_{j,g}^{n+1,s} ,\; \; \left(\sigma _{a,g} \right)_{j,g}^{n+1,s} ,\; \; \left(D^{1} \right)_{j,k,g}^{n+1,s} ,\; \; \left(D^{2} \right)_{j,k,g}^{n+1,s} $ with $T_{j}^{n+1,s} $.

2.2) Solve the following equations to find the $\rho _{j,g}^{n+1,s+1} $ and $T_{j}^{n+1,s+1} $.
\begin{equation} \label{Equation_3_25_}
\left\{\begin{array}{l} {\rho _{j,g}^{n+1,s+1} =\rho _{j,g}^{n} -\frac{\Delta t}{V_{j} } \sum _{k}\Phi _{j,k,g}^{n+1,s}  +\frac{c\Delta tL_{a}^{\varepsilon } }{\varepsilon } \left(2\pi \left(\sigma _{e,g} \right)_{j}^{n+1,s} \phi _{j,g}^{n+1,s+1} -\left(\sigma _{a,g} \right)_{j}^{n+1,s} \rho _{j,g}^{n+1,s+1} \right)} \\ {C_{V} \left(T_{j}^{n+1,s+1} -T_{j}^{n} \right)=\frac{\Delta tL_{a}^{\varepsilon } }{\varepsilon } \sum _{g}\left(\left(\sigma _{a,g} \right)_{j}^{n+1,s} \rho _{j,g}^{n+1,s+1} -2\pi \left(\sigma _{e,g} \right)_{j}^{n+1,s} \phi _{j,g}^{n+1,s+1} \right) } \\ {\phi _{j,g}^{n+1,s+1} =\phi _{j,g}^{n+1,s} +\left(\frac{\partial \phi _{g} }{\partial T} \right)_{j}^{n+1,s} \left(T_{j}^{n+1,s+1} -T_{j}^{n+1,s} \right)} \\ {\frac{\partial \phi _{g} }{\partial T} =\int _{\nu _{g-{1\mathord{\left/ {\vphantom {1 2}} \right. \kern-\nulldelimiterspace} 2} } }^{\nu _{g+{1\mathord{\left/ {\vphantom {1 2}} \right. \kern-\nulldelimiterspace} 2} } }\frac{\partial B\left(\nu ,T\right)}{\partial T} d\nu  } \end{array}\right.
\end{equation}
where $\left({\partial \phi _{g} \mathord{\left/ {\vphantom {\partial \phi _{g}  \partial T}} \right. \kern-\nulldelimiterspace} \partial T} \right)_{j}^{n+1,s} $ is a function of $T_{j}^{n+1,s} $, and the macroscopic flux $\Phi _{j,k,g}^{n+1,s} $ has the same form as \eqref{Equation_3_22_}, which can be written as:
\begin{equation} \label{Equation_3_26_}
\begin{array}{l} {\Phi _{j,k,g}^{n+1,s} =\left\langle \frac{cl_{k} }{\varepsilon \Delta t} \int _{t_{n} }^{t_{n+1} }\mu 'e^{-\left(\lambda _{g} \right)_{j,k} \left(t-t_{n} \right)} I_{g} \left(t_{n} ,-\frac{\mu 'c}{\varepsilon } \left(t-t_{n} \right),-\frac{\xi 'c}{\varepsilon } \left(t-t_{n} \right)\right)dt \right\rangle } \\ {+\frac{2\pi }{3} \left(D^{1} \right)_{j,k,g}^{n+1,s} \delta _{x'} \phi _{j,k,g}^{n+1,s+1} +\frac{2\pi }{3} \left(D^{2} \right)_{j,k,g}^{n+1,s} \delta _{x'} \rho _{j,k,g}^{n+1,s+1} } \end{array}
\end{equation}

2.3) Compute the relative iteration error. Stop the iteration when convergent condition is reached.

\begin{enumerate}
\item  Update the solutions $\rho _{j,g}^{n+1} =\rho _{j,g}^{n+1,s+1} $ and $T_{j}^{n+1} =T_{j}^{n+1,s+1} $.
\end{enumerate}

With solved $T_{j}^{n+1} $ above, the $\phi _{j,g}^{n+1} $ could be calculated through \eqref{Equation_3_4_}. Up to now, the macroscopic terms in \eqref{Equation_3_16_} and \eqref{Equation_3_22_} have been solved.

\subsection{Microscopic solver for the radiation intensity}
In this subsection, we will show how to determine the microscopic terms in \eqref{Equation_3_22_} by the particle-based Monte Carlo method, following the methodology of UGKP.

With the solved macroscopic quantities $\rho _{j,g}^{n+1} $ and $\phi _{j,g}^{n+1} $ above, we could reconstruct the macroscopic terms of radiation intensity at $t_{n+1} $ based on equation \eqref{Equation_3_16_}. Under the first order approximation of \eqref{Equation_3_27_},
\begin{equation} \label{Equation_3_27_}
\phi _{g} \left(s,-\frac{\mu 'c}{\varepsilon } \left(t_{n+1} -s\right),-\frac{\xi 'c}{\varepsilon } \left(t_{n+1} -s\right)\right)=\phi _{j,g}^{n+1} ,\; \rho _{g} \left(s,-\frac{\mu 'c}{\varepsilon } \left(t_{n+1} -s\right),-\frac{\xi 'c}{\varepsilon } \left(t_{n+1} -s\right)\right)=\rho _{j,g}^{n+1}
\end{equation}
we have the macroscopic terms in following form:
\begin{equation} \label{Equation_3_28_}
\begin{array}{l} {\int _{t_{n} }^{t_{n+1} }e^{-\lambda _{g} \left(t_{n+1} -s\right)} \left[\begin{array}{l} {\frac{cL_{a}^{\varepsilon } \sigma _{e,g} }{\varepsilon } \phi _{g} \left(s,-\frac{\mu 'c}{\varepsilon } \left(t_{n+1} -s\right),-\frac{\xi 'c}{\varepsilon } \left(t_{n+1} -s\right)\right)} \\ {+\frac{cL_{s}^{\varepsilon } \sigma _{s,g} }{\varepsilon } \frac{\rho _{g} }{{\rm 2}\pi } \left(s,-\frac{\mu 'c}{\varepsilon } \left(t_{n+1} -s\right),-\frac{\xi 'c}{\varepsilon } \left(t_{n+1} -s\right)\right)} \end{array}\right]ds } \\ {=\frac{1}{\lambda _{g} } \left(1-e^{-\lambda _{g} \Delta t} \right)\left[\frac{cL_{a}^{\varepsilon } \sigma _{e,g} }{\varepsilon } \phi _{j,g}^{n+1} +\frac{cL_{s}^{\varepsilon } \sigma _{s,g} }{\varepsilon } \frac{\rho _{j,g}^{n+1} }{{\rm 2}\pi } \right]} \end{array}
\end{equation}
Thus, we could employ the Monte Carlo method to solve the radiation intensity with equation \eqref{Equation_3_16_} and \eqref{Equation_3_28_}.

Based on \eqref{Equation_3_16_} and \eqref{Equation_3_28_}, there are three types of sources: the source from previous time step $I_{g} \left(t_{n} \right)$, the emission source $\phi _{j,g}^{n+1} $, and the scattering source ${\rho _{j,g}^{n+1} \mathord{\left/ {\vphantom {\rho _{j,g}^{n+1}  {\rm 2}\pi }} \right. \kern-\nulldelimiterspace} {\rm 2}\pi } $. The factor $e^{-\lambda _{g} \Delta t} $ with $I_{g} \left(t_{n} \right)$ is the probability of photon to free stream without collision, while the same probability for the emission source $\phi _{j,g}^{n+1} $ and scattering source ${\rho _{j,g}^{n+1} \mathord{\left/ {\vphantom {\rho _{j,g}^{n+1}  {\rm 2}\pi }} \right. \kern-\nulldelimiterspace} {\rm 2}\pi } $ is $\left(1-e^{-\lambda _{g} \Delta t} \right)$, coming from the integral of exponential function from $t_{n} $ to $t_{n+1} $. In addition, the factor ${cL_{a}^{\varepsilon } \sigma _{e,g} \mathord{\left/ {\vphantom {cL_{a}^{\varepsilon } \sigma _{e,g}  \lambda _{g} \varepsilon }} \right. \kern-\nulldelimiterspace} \lambda _{g} \varepsilon } $ and ${cL_{s}^{\varepsilon } \sigma _{s,g} \mathord{\left/ {\vphantom {cL_{s}^{\varepsilon } \sigma _{s,g}  \lambda _{g} \varepsilon }} \right. \kern-\nulldelimiterspace} \lambda _{g} \varepsilon } $ indicate the relative contribution of emission and scattering in the macroscopic source, depended on the cross sections.

Unlike Monte Carlo method for neutron transport which utilizes a dimensionless weight, an energy weight is employed here to make sure the conservation of energy. With above discussion, we could sample photons from different sources. First, for the source from previous time step, it contains photons from the boundary condition, the initial condition, and/or the free transport photons from the previous time step. Since the free transport photons from the previous time step are already known, we only need to sample photons from the boundary condition and initial condition. Then, for the emission source, we sample the photons as angular isotropic and spatial uniformly in each cell. For the last source of scattering, the sampling strategy is the same as the emission source above.

With sampled photons above and those existing census photons in the domain, we need to track the trajectory of each photon during the next time step from $t_{n+1} $ to $t_{n+2} $. The process of photon tracking in each time step is following. There are essentially three events that should be considered during each time step: (i) the photon may collide with an atom and then either be absorbed or be scattered; (ii) the photon may exit the current cell and enter a cell with different opacity or leak out of the system; (iii) the photon may survive at the end of time step and go to census. Associated with the three events are three distances -- the distance to collision $d_{C} $, the distance to the cell interface $d_{B} $, and the distance that would be traveled until the end of time step $d_{T} $. The calculation of each quantity is straightforward. The distance to collision $d_{C} $ is sampled by:
\begin{equation} \label{Equation_3_29_}
d_{C} =-\frac{1}{\Sigma _{t} } \ln \chi
\end{equation}
where $\Sigma _{t} $ is the macroscopic total cross section of the medium in current cell, and $\chi $ is a random number on $\left[0,1\right]$. The distance to the spatial cell interface $d_{B} $ satisfies:
\begin{equation} \label{Equation_3_30_}
\vec{r}_{B} -\vec{r}=d_{B} \vec{\Omega }
\end{equation}
where $\vec{r}$ is the original location of each photon and is $\vec{r}_{B} $ the cell interface location in direction $\vec{\Omega }$. And the distance travelled to the end of time step is
\begin{equation} \label{Equation_3_31_}
d_{t} =c\left(t_{n+2} -t\right)
\end{equation}
where $t$ is the current time of each particle. With calculated distances in \eqref{Equation_3_29_} - \eqref{Equation_3_31_}, the particle-based Monte Carlo solver will determine which event should happen, based on the minimum quantity of these three distances. If the photon has collided with an atom, gone to census, or leaked out of the system, then the photon tracking process ends for current photon in this step.

During the evolution of photon trajectory, the only one quantity should be tallied is the net flux of the photons in each cell surface, which can be calculated by:
\begin{equation} \label{Equation_3_32_}
\Phi _{j,k,g}^{micro} =\sum _{i}sign\left(\Omega _{x'} \right)w_{i,g}
\end{equation}
where \textit{i} denotes the photon that transport across the edge \textit{k}. This quantity provides the microscopic term in \eqref{Equation_3_22_}, which are needed to solve the macroscopic equations \eqref{Equation_3_10_}, as discussed in subsection 3.4. Thus, this particle-based Monte Carlo completes the construction of the UGKP for the frequency-dependent radiation transfer equations \eqref{Equation_2_1_}.

\subsection{Summary}
In the previous subsections, we have introduced the macro and microscopic solvers for the UGKP method for frequency-dependent radiative transfer system. Here we will give a summary of the overall UGKP algorithm. The flowchart for the computation procedures is given in Figure 2 for better understand of the algorithm.

For each time step, the procedure for UGKP method consists of the following steps:

Step 1. Sample the Monte Carlo particles from the source from previous time step, the emission source, and the scattering source;

Step 2. Track the trajectories of these photons and those existing census photons in the domain, and tally the net flux across each cell interface;

Step 3. Solve the macroscopic system based on the tallied fluxes to obtain the macroscopic quantities $\rho _{j,g}^{n+1} $, $\phi _{j,g}^{n+1} $ and $T_{j}^{n+1} $.
\begin{figure}
  \centering
  \includegraphics[width=0.6\textwidth]{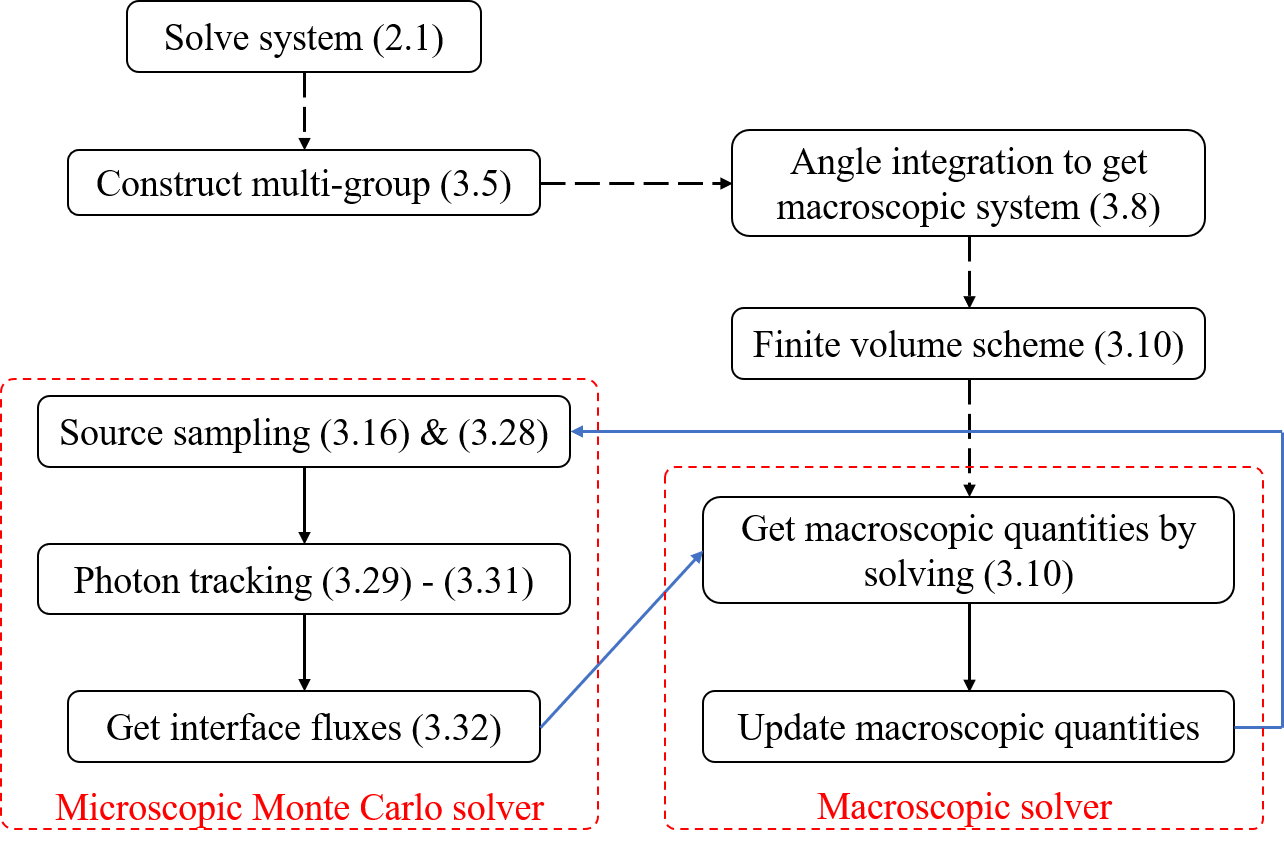}
  \caption{The flowchart of UGKP.}
  \label{fig_image2}
\end{figure}

In this UGKP algorithm, the evolution of microscopic particles provides a part of the numerical flux for the macroscopic equations, and the solution of macroscopic equations provides the sources for the microscopic solver. The macro and microscopic quantities are tightly coupled and updated together.

\section{Asymptotic analysis}\label{analysis}

For radiative transfer methods, to have the asymptotic preserving (AP) property is important for any scheme. In this section, we will prove that the extended UGKP method has the AP property in both the free transport and different equilibrium states in the diffusive regime.

\subsection{Free transport regime}

In the free transport limit, both $\sigma _{a} $ and $\sigma _{s} $ tend to 0 while the parameter $\varepsilon $, $L_{a}^{\varepsilon } $ and $L_{s}^{\varepsilon } $ is fixed as constant (since $L_{a}^{\varepsilon } $ and $L_{a}^{\varepsilon } $ are depended on $\varepsilon $). For simplicity, we take the following assumption:
\begin{equation} \label{Equation_4_1_}
\sigma _{a} =\sigma _{s} =\sigma \to 0,\; \; L_{a}^{\varepsilon } =L_{s}^{\varepsilon } =\varepsilon \to conest
\end{equation}
Thus, the coefficients in the macroscopic numerical flux \eqref{Equation_3_26_} satisfy
\begin{equation} \label{Equation_4_2_}
\begin{array}{l} {\sigma _{e,g} \to o\left(\sigma \right)\to 0,\; \; \sigma _{a,g} \to o\left(\sigma \right)\to 0,\; \; \sigma _{s,g} \to o\left(\sigma \right)\to 0,} \\ {\lambda _{g} =c\left(\sigma _{a,g} +\sigma _{s,g} \right)\to o\left(\sigma \right)\to {\rm 0,}\; \; e^{-\lambda _{g} \left(t-t_{n} \right)} \to 1} \end{array}
\end{equation}
and
\begin{equation} \label{Equation_4_3_}
\begin{array}{l} {\left(D^{1} \right)_{j,k,g}^{n+1} \to -\frac{2c\left(\sigma _{e,g} \right)_{j,k,g}^{n+1} l_{k} }{\varepsilon ^{{\rm 2}} \left[\left(\sigma _{a,g} \right)_{j,k,g}^{n+1} +\left(\sigma _{s,g} \right)_{j,k,g}^{n+1} \right]^{{\rm 2}} } \to o\left(\sigma ^{-1} \right){\rm ,}} \\ {\left(D^{2} \right)_{j,k,g}^{n+1} \to -\frac{2c\left(\sigma _{s,g} \right)_{j,k,g}^{n+1} l_{k} }{2\pi \varepsilon ^{{\rm 2}} \left[\left(\sigma _{a,g} \right)_{j,k,g}^{n+1} +\left(\sigma _{s,g} \right)_{j,k,g}^{n+1} \right]^{{\rm 2}} } \to o\left(\sigma ^{-1} \right)} \end{array}
\end{equation}
With above equations, we substitute the macroscopic numerical flux \eqref{Equation_3_26_} into macroscopic equation \eqref{Equation_3_10_} and compare terms that are the same order in $\sigma $.

The $o\left(\sigma ^{-1} \right)$ equation is
\begin{equation} \label{Equation_4_4_}
\sum _{k}\frac{\phi _{j',g}^{n+1} -\phi _{j,g}^{n+1} -\left(\tau _{j,k}^{-} +\tau _{j,k}^{+} \right)\left(\hat{\phi }_{k,2}^{n+1} -\hat{\phi }_{k,1}^{n+1} \right)}{l_{j,k}^{-} +l_{j,k}^{+} }  +\sum _{k}\frac{\rho _{j',g}^{n+1} -\rho _{j,g}^{n+1} -\left(\tau _{j,k}^{-} +\tau _{j,k}^{+} \right)\left(\hat{\rho }_{k,2}^{n+1} -\hat{\rho }_{k,1}^{n+1} \right)}{l_{j,k}^{-} +l_{j,k}^{+} } =0
\end{equation}
with a solution as
\begin{equation} \label{Equation_4_5_}
\phi _{j,g}^{n+1} =\phi _{g}^{n+1} ,\; \; \rho _{j,g}^{n+1} =\rho _{g}^{n+1}
\end{equation}
This means that the macroscopic quantity $\phi _{g}^{n+1} $ in all cells is the same, as well as the macroscopic quantity $\rho _{g}^{n+1} $. Thus, the macroscopic numerical flux \eqref{Equation_3_26_} has the following form:
\begin{equation} \label{Equation_4_6_}
\Phi _{j,k,g}^{n+1} \to \left\langle \frac{cl_{k} }{\varepsilon \Delta t} \int _{t_{n} }^{t_{n+1} }\mu 'I_{g} \left(t_{n} ,-\frac{\mu 'c}{\varepsilon } \left(t-t_{n} \right),-\frac{\xi 'c}{\varepsilon } \left(t-t_{n} \right)\right)dt \right\rangle
\end{equation}
This equation indicates that the macroscopic numerical flux is exactly the same as the tallied free transport flux of the microscopic Monte Carlo particles.

The $o\left(\sigma ^{{\rm 0}} \right)$ equation is
\begin{equation} \label{Equation_4_7_}
\rho _{i,j,g}^{n{\rm +1}} =\rho _{i,j,g}^{n} -\frac{\Delta t}{V_{j} } \sum _{k}\left\langle \frac{cl_{k} }{\varepsilon \Delta t} \int _{t_{n} }^{t_{n+1} }\mu 'I_{g} \left(t_{n} ,-\frac{\mu 'c}{\varepsilon } \left(t-t_{n} \right),-\frac{\xi 'c}{\varepsilon } \left(t-t_{n} \right)\right)dt \right\rangle
\end{equation}
which indicates that the radiation energy changes only because the contribution of the tallied free transport flux of the microscopic Monte Carlo particles.

Moreover, the $o\left(\sigma ^{1} \right)$ equation is
\begin{equation} \label{Equation_4_8_}
2\pi \phi _{i,j,g}^{n+1} -\rho _{i,j,g}^{n+1} =0
\end{equation}
which can be further deducted:
\begin{equation} \label{Equation_4_9_}
\sum _{g}2\pi \phi _{i,j,g}^{n+1}  =\sum _{g}\rho _{i,j,g}^{n+1}  \Rightarrow \int _{0}^{\infty }\int _{2\pi }B\left(\nu ,T\right)d\vec{\Omega }d\nu   =\int _{0}^{\infty }\int _{2\pi }Id\vec{\Omega }d\nu   \Rightarrow I=B\left(\nu ,T\right)
\end{equation}
This means that the system has reach the equilibrium state of free transport. Above results confirmed that the solution of the extended UGKP method could recover the free transport solution and be asymptotic preserving in the free transport limit.

\subsection{Equilibrium diffusion regime}
For the equilibrium diffusion limit, both $\sigma _{a} $ and $\sigma _{s} $ are positive, while the parameter $\varepsilon $ tends to 0 with $L_{a}^{\varepsilon } ={1\mathord{\left/ {\vphantom {1 \varepsilon }} \right. \kern-\nulldelimiterspace} \varepsilon } $ and $L_{s}^{\varepsilon } =\varepsilon $. Thus, the coefficients in the macroscopic numerical flux \eqref{Equation_3_26_} satisfy
\begin{equation} \label{Equation_4_10_}
\lambda _{g} =\frac{c\left({\sigma _{a,g} \mathord{\left/ {\vphantom {\sigma _{a,g}  \varepsilon }} \right. \kern-\nulldelimiterspace} \varepsilon } +\varepsilon \sigma _{s,g} \right)}{\varepsilon } \to o\left(\varepsilon ^{-2} \right){\rm ,}\; \; e^{-\lambda _{g} \left(t-t_{n} \right)} \to 0
\end{equation}
and
\begin{equation} \label{Equation_4_11_}
\begin{array}{l} {\left(D^{1} \right)_{j,k,g}^{n+1} \to -\frac{c\left(\sigma _{e,g} \right)_{j,k,g}^{n+1} l_{k} }{\varepsilon ^{{\rm 2}} \left[{\left(\sigma _{a,g} \right)_{j,k,g}^{n+1} \mathord{\left/ {\vphantom {\left(\sigma _{a,g} \right)_{j,k,g}^{n+1}  \varepsilon }} \right. \kern-\nulldelimiterspace} \varepsilon } +\varepsilon \left(\sigma _{s,g} \right)_{j,k,g}^{n+1} \right]^{{\rm 2}} } \to -\frac{c\left(\sigma _{e,g} \right)_{j,k,g}^{n+1} l_{k} }{\left[\left(\sigma _{a,g} \right)_{j,k,g}^{n+1} \right]^{2} } {\rm ,}} \\ {\left(D^{2} \right)_{j,k,g}^{n+1} \to -\frac{c\left(\sigma _{s,g} \right)_{j,k,g}^{n+1} l_{k} }{2\pi \left[{\left(\sigma _{a,g} \right)_{j,k,g}^{n+1} \mathord{\left/ {\vphantom {\left(\sigma _{a,g} \right)_{j,k,g}^{n+1}  \varepsilon }} \right. \kern-\nulldelimiterspace} \varepsilon } +\varepsilon \left(\sigma _{s,g} \right)_{j,k,g}^{n+1} \right]^{{\rm 2}} } \to o\left(\varepsilon ^{2} \right)\to 0} \end{array}
\end{equation}
Thus, the macroscopic numerical flux \eqref{Equation_3_26_} has the following form:
\begin{equation} \label{Equation_4_12_}
\Phi _{j,k,g}^{n+1} \to -\frac{2\pi }{3} \frac{c\left(\sigma _{e,g} \right)_{j,k,g}^{n+1} l_{k} }{\left[\left(\sigma _{a,g} \right)_{j,k,g}^{n+1} \right]^{2} } \delta _{x'} \phi _{j,k,g}^{n+1}
\end{equation}
Then we substitute the macroscopic numerical flux \eqref{Equation_4_12_} into the macroscopic equations \eqref{Equation_3_10_} and compare terms that are the same order in $\varepsilon $.

The $o\left(\varepsilon ^{-{\rm 2}} \right)$ equation is
\begin{equation} \label{Equation_4_13_}
2\pi \left(\sigma _{e,g} \right)_{i,j,g}^{n+1} \phi _{i,j,g}^{n+1} -\left(\sigma _{a,g} \right)_{i,j,g}^{n+1} \rho _{i,j,g}^{n+1} =0
\end{equation}
Summing the above equation over the group index \textit{g}, and by the definitions of $\sigma _{e,g} $ and $\sigma _{a,g} $ in \eqref{Equation_3_3_}, we get
\begin{equation} \label{Equation_4_15_}
\begin{array}{l} {\int _{{\rm 0}}^{\infty }\int _{2\pi }\sigma _{a} B\left(\nu ,T\right)d\vec{\Omega }d\nu   =\int _{{\rm 0}}^{\infty }\int _{2\pi }\sigma _{a} Id\vec{\Omega }d\nu   \Rightarrow I=B\left(\nu ,T\right)} \\ {\Rightarrow 2\pi \phi _{i,j,g}^{n+1} =\rho _{i,j,g}^{n+1} ,\; \; \left(\sigma _{e,g} \right)_{i,j,g}^{n+1} =\left(\sigma _{a,g} \right)_{i,j,g}^{n+1} =\left(\sigma _{P,g} \right)_{i,j,g}^{n+1} } \end{array}
\end{equation}
which indicates that the leading order radiation temperature approaches to the material temperature at the equilibrium limit.

Summing the macroscopic radiation transport equation over the group index \textit{g }and combining it with the material energy equation in \eqref{Equation_3_10_}, we have the $o\left(\varepsilon ^{{\rm 0}} \right)$ equation as
\begin{equation} \label{Equation_4_16_}
\begin{array}{l} {\sum _{g=1}^{G}\rho _{j,g}^{n{\rm +1}}  =\sum _{g=1}^{G}\rho _{j,g}^{n}  -\frac{\Delta t}{V_{j} } \sum _{g=1}^{G}\sum _{k}\left(-\frac{2\pi }{3} \frac{cl_{k} }{\left(\sigma _{P,g} \right)_{j,k,g}^{n+1} } \delta _{x'} \phi _{j,k,g}^{n+1} \right)  } \\ {-cC_{V} \left(T_{i,j}^{n+1} -T_{i,j}^{n} \right)} \end{array}
\end{equation}
Due to the relations \eqref{Equation_4_15_} and \eqref{Equation_3_4_}, we get
\begin{equation} \label{Equation_4_17_}
\left\{\begin{array}{l} {\sum _{g}2\pi \phi _{i,j,g}^{n+1}  =\int _{{\rm 0}}^{\infty }\int _{2\pi }\sigma _{a} B\left(\nu ,T_{i,j}^{n+1} \right)d\vec{\Omega }d\nu   =ac\left(T_{i,j}^{n+1} \right)^{4} } \\ {\sum _{g}\rho _{i,j,g}^{n+1}  =\sum _{g}2\pi \phi _{i,j,g}^{n+1}  =ac\left(T_{i,j}^{n+1} \right)^{4} } \end{array}\right.
\end{equation}
As a result, equation \eqref{Equation_4_16_} becomes a standard nine points scheme for the diffusion limit equation \eqref{Equation_2_3_} combined with \eqref{Equation_4_17_}. It can be further reduced to the standard five points scheme for orthogonal quadrilateral mesh. This shows that the extended UGKP method for the frequency-dependent radiative transfer equations \eqref{Equation_2_1_} is an asymptotic preserving (AP) method in the equilibrium diffusion limit.

\subsection{Non-equilibrium diffusion regime}
For the non-equilibrium diffusion limit, both $\sigma _{a} $ and $\sigma _{s} $ are positive, while the parameter $\varepsilon $ tends to 0 with $L_{a}^{\varepsilon } =\varepsilon $ and $L_{s}^{\varepsilon } ={{\rm 1}\mathord{\left/ {\vphantom {{\rm 1} \varepsilon }} \right. \kern-\nulldelimiterspace} \varepsilon } $. Thus, the coefficients in the macroscopic numerical flux \eqref{Equation_3_26_} satisfy
\begin{equation} \label{Equation_4_17_2}
\lambda _{g} =\frac{c\left(\varepsilon \sigma _{a,g} +{\sigma _{s,g} \mathord{\left/ {\vphantom {\sigma _{s,g}  \varepsilon }} \right. \kern-\nulldelimiterspace} \varepsilon } \right)}{\varepsilon } \to o\left(\varepsilon ^{-2} \right){\rm ,}\; \; e^{-\lambda _{g} \left(t-t_{n} \right)} \to 0
\end{equation}
and
\begin{equation} \label{Equation_4_18_}
\begin{array}{l} {\left(D^{1} \right)_{j,k,g}^{n+1} \to -\frac{c\left(\sigma _{e,g} \right)_{j,k,g}^{n+1} l_{k} }{\left[\varepsilon \left(\sigma _{a,g} \right)_{j,k,g}^{n+1} +{\left(\sigma _{s,g} \right)_{j,k,g}^{n+1} \mathord{\left/ {\vphantom {\left(\sigma _{s,g} \right)_{j,k,g}^{n+1}  \varepsilon }} \right. \kern-\nulldelimiterspace} \varepsilon } \right]^{{\rm 2}} } \to o\left(\varepsilon ^{2} \right)\to 0{\rm ,}} \\ {\left(D^{2} \right)_{j,k,g}^{n+1} \to -\frac{c\left(\sigma _{s,g} \right)_{j,k,g}^{n+1} l_{k} }{2\pi \varepsilon ^{{\rm 2}} \left[\varepsilon \left(\sigma _{a,g} \right)_{j,k,g}^{n+1} +{\left(\sigma _{s,g} \right)_{j,k,g}^{n+1} \mathord{\left/ {\vphantom {\left(\sigma _{s,g} \right)_{j,k,g}^{n+1}  \varepsilon }} \right. \kern-\nulldelimiterspace} \varepsilon } \right]^{{\rm 2}} } \to -\frac{cl_{k} }{2\pi \left(\sigma _{s,g} \right)_{j,k,g}^{n+1} } } \end{array}
\end{equation}
Thus, the macroscopic numerical flux \eqref{Equation_3_26_} has the following form:
\begin{equation} \label{Equation_4_19_}
\Phi _{j,k,g}^{n+1} \to -\frac{cl_{k} }{3\left(\sigma _{s,g} \right)_{j,k,g}^{n+1} } \delta _{x'} \rho _{j,k,g}^{n+1}
\end{equation}
Substitute the macroscopic numerical flux \eqref{Equation_4_19_} into the macroscopic equations \eqref{Equation_3_10_}, summing the macroscopic radiation transport equation over the group index \textit{g }and combining it with the material energy equation, we have:
\begin{equation} \label{Equation_4_20_}
\begin{array}{l} {\sum _{g=1}^{G}\rho _{j,g}^{n{\rm +1}}  =\sum _{g=1}^{G}\rho _{j,g}^{n}  -\frac{\Delta t}{V_{j} } \sum _{g=1}^{G}\sum _{k}\left(-\frac{cl_{k} }{3\left(\sigma _{s,g} \right)_{j,k,g}^{n+1} } \delta _{x'} \rho _{j,k,g}^{n+1} \right)  } \\ {+c\Delta t\sum _{g=1}^{G}\left(2\pi \left(\sigma _{e,g} \right)_{j}^{n+1} \phi _{j,g}^{n{\rm +1}} -\left(\sigma _{a,g} \right)_{j}^{n+1} \rho _{j,g}^{n{\rm +1}} \right) } \end{array}
\end{equation}

Equation \eqref{Equation_4_20_} is a standard nine points scheme for the first diffusion limit equation in \eqref{Equation_2_4_}, which is independent of the parameter $\varepsilon $. And the second equation of \eqref{Equation_2_4_} is also independent of the parameter $\varepsilon $. Thus, the convergence of equations \eqref{Equation_2_4_} is automatically satisfied. This indicates that the extended UGKP method for the frequency-dependent radiative transfer equations \eqref{Equation_2_1_} is also an asymptotic preserving (AP) method in the non-equilibrium diffusion limit.

\section{Numerical examples}\label{examples}
In this section, we present six numerical examples to validate the extended UGKP method. These examples include two one-dimensional Marshak wave problems, the tophat problem, a multi-group problem, and two modified Marshak wave problems. In the following computations, the unit of length is taken to be centimeter (cm), mass unit is gramme (g), time unit is nanosecond (ns), temperature unit is kilo electronvolt (keV), and energy unit is 10${}^{9}$ Joules (GJ). Under the above units, the speed of light is 29.98 cm/ns, and the radiation constant \textit{a} is 0.01372 GJ/(cm${}^{3}$keV${}^{4}$).

\subsection{Marshak wave-2A problem}
First, we take the one-dimensional Marshak wave problems to test the extended UGKP method. In the Marshak wave-2A problem, a thermal wave was driven by a constant intensity incident on the left boundary of the computational domain. The temperature of the left boundary source is 1 keV, while the initial material and radiation temperature is in equilibrium at 10${}^{-6}$ keV. The computational domain is a slab of 1.0 cm thick which consists of unstructured mesh with maximum size of 0.005 cm. The absorption/emission coefficient is set to be temperature-dependent of $\sigma ={30.0\mathord{\left/ {\vphantom {30.0 T^{3} }} \right. \kern-\nulldelimiterspace} T^{3} } cm^{-1} $, and the specific heat to be 0.3 GJ/keV/cm${}^{3}$.

In Figure 3 (a), the material and radiation temperatures simulated using UGKP method at times 0.2, 0.4, 0.6, 0.8, 1.0 ns are given, which is the same as the results in \cite{shi2020asymptotic}. The small absorption/emission coefficient violates the equilibrium diffusion approximation in this case. Thus, the computed UGKP material temperatures are quite different from the diffusion equations results in Figure 3 (b). These results show that the AP scheme works well for this problem on the one hand, and the extended UGKP method could degenerate into the gray radiative transfer equation on the other hand.

\begin{figure}
  \centering
  \subfigure[]{\includegraphics[width=0.49\textwidth]{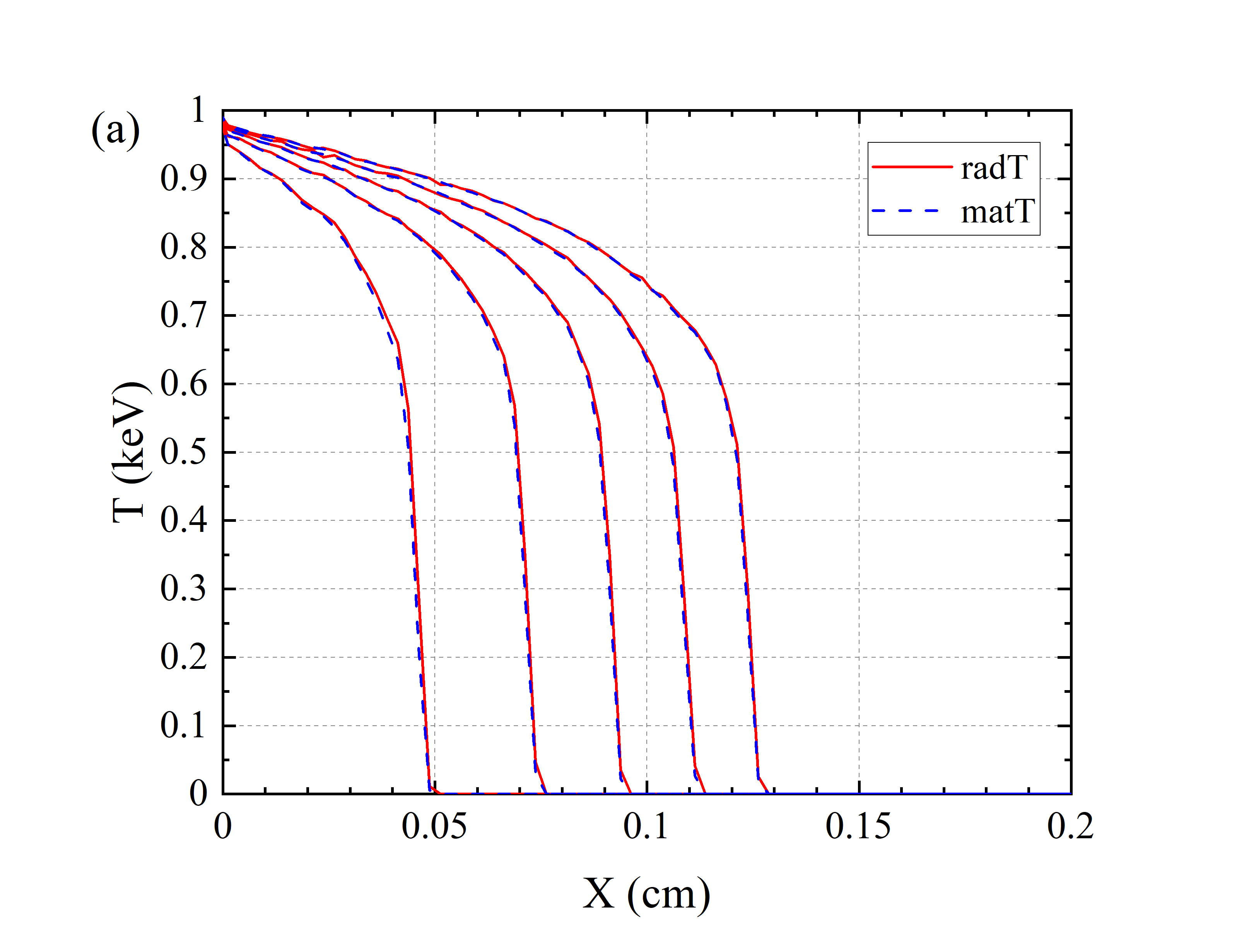}}
  \subfigure[]{\includegraphics[width=0.49\textwidth]{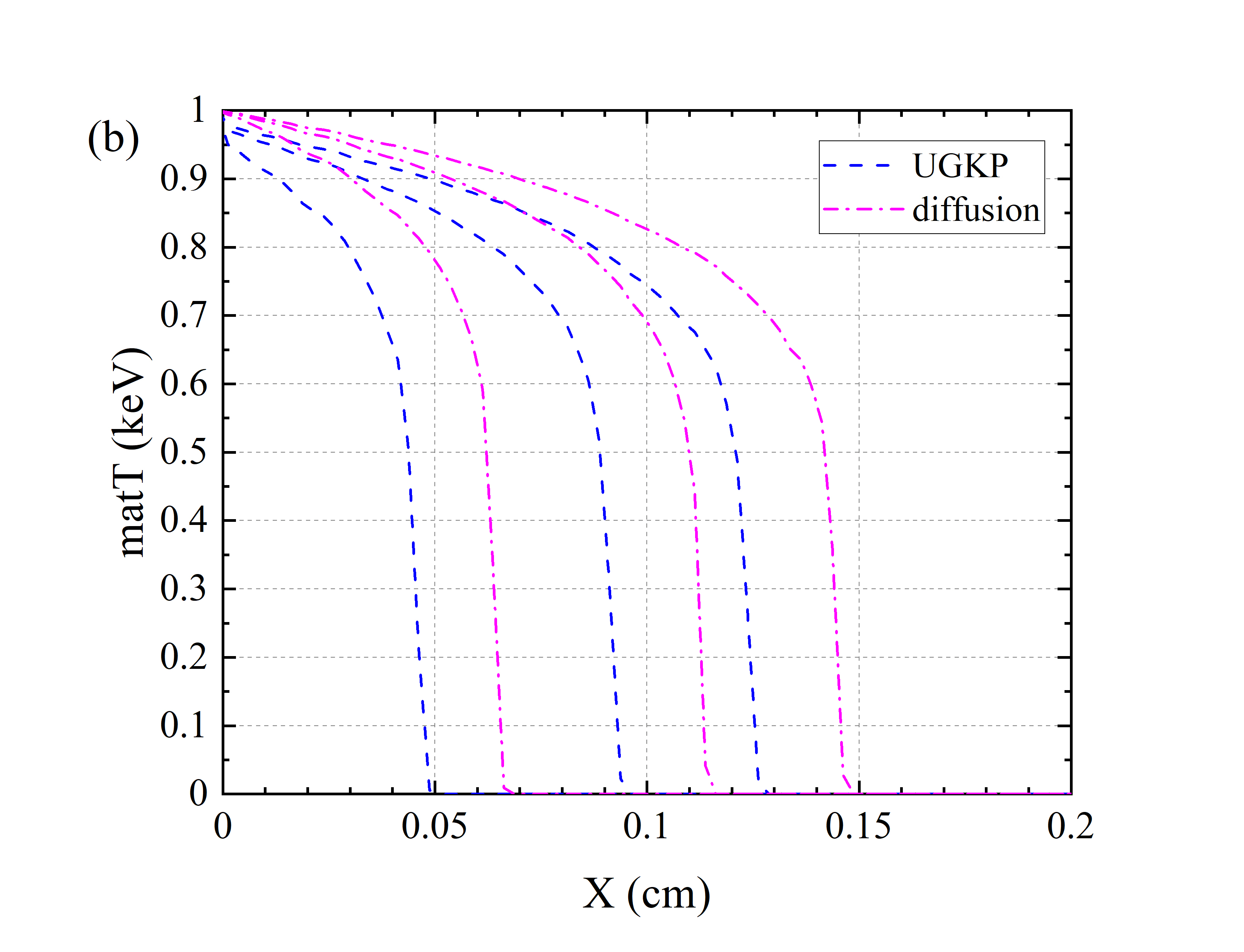}}
  \caption{The results of Marshak wave-2A problem. (a) The radiation and material temperature at times 0.2, 0.4, 0.6, 0.8, 1.0 ns respectively. (b) The material temperature from UGKP simulation and the diffusion equation solution at time 0.2, 0.6, 1.0 ns respectively.}
  \label{fig_image3}
\end{figure}

\subsection{Marshak wave-2B problem}
The Marshak wave-2B problem is exactly the same as the Marshak wave-2A problem except that it has a temperature-dependent absorption/emission coefficient with $\sigma ={30{\rm 0}.0\mathord{\left/ {\vphantom {30{\rm 0}.0 T^{3} }} \right. \kern-\nulldelimiterspace} T^{3} } cm^{-1} $. With this absorption/emission coefficient, the solution will get the equilibrium diffusion limit solution.

The material and radiation temperatures simulated using UGKP method at times 15, 30, 45, 60, 74 ns are given in Figure 4 (a), which is also the same as the results in [31]. In Figure 4 (b), the computed material temperatures for both UGKP simulation and diffusion equations results at times 15, 45, 74 ns are given. With this large absorption/emission coefficient case, the UGKP results are closer to the diffusive limit results than Marshak wave-2A problem.
\begin{figure}
  \centering
  \subfigure[]{\includegraphics[width=0.49\textwidth]{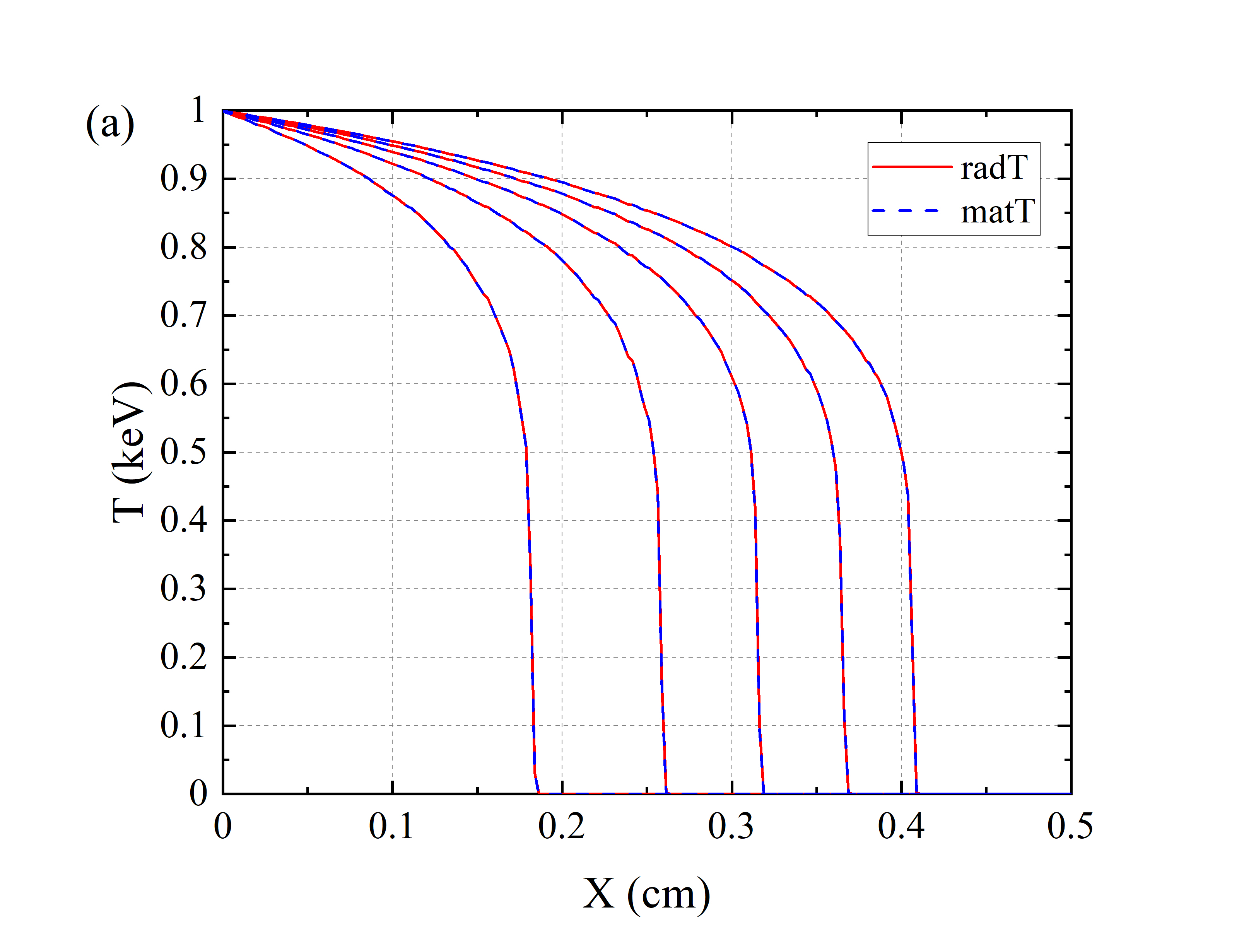}}
  \subfigure[]{\includegraphics[width=0.49\textwidth]{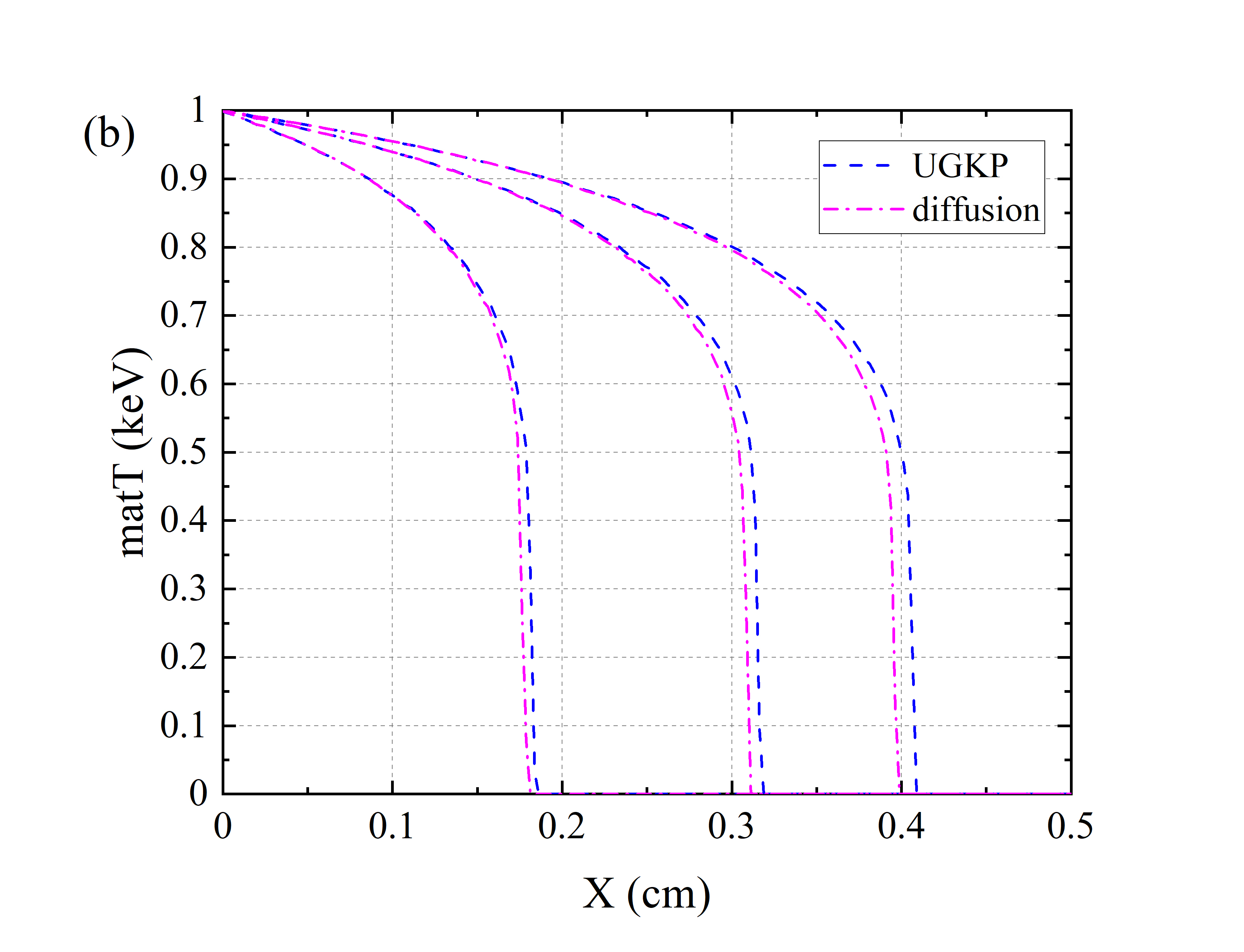}}
  \caption{The results of Marshak wave-2B problem. (a) The radiation and material temperature at times 15, 30, 45, 60, 74 ns respectively. (b) The material temperature from UGKP simulation and the diffusion equation solution at time 15, 45, 74 ns respectively.}
  \label{fig_image5}
\end{figure}

\subsection{Tophat problem}
Beside the Marshak wave problems, we also consider the Tophat problem. The Tophat problem, also known as the crooked pipe problem, is a two-dimensional problem consists of both optically thin and thick material. The original definition was in cylindrical coordinates, while we solve this problem in Cartesian coordinates. The size of the computational domain is [0,7]$\mathrm{\times}$[-2,2], which consists of unstructured mesh with maximum size of 0.08 cm. The dense, opaque material with density 10 g/cm${}^{3}$ and opacity $\sigma ={\rm 2000}\, cm^{-1} $ is located in the regions [3,4]$\mathrm{\times}$[-1,1], [0,2.5]$\mathrm{\times}$[0.5,2], [0,2.5]$\mathrm{\times}$[-2,-0.5], [4.5,7]$\mathrm{\times}$[0.5,2], [4.5,7]$\mathrm{\times}$[-2,-0.5], [2.5,4.5]$\mathrm{\times}$[1.5,2], [2.5,4.5]$\mathrm{\times}$[-2,-1.5]. And the optical thin material with density 0.01 g/cm${}^{3}$ and opacity $\sigma ={\rm 0.2}\, cm^{-1} $ occupies the other regions, which forming the pipe. The heat capacity is 0.1 GJ/g/keV for both optically thin and thick materials. Initially, the system is in equilibrium at 0.05 keV, while a surface source at 0.5 keV is located on the left boundary of the pipe for $-0.5<y<0.5$. In addition, five probes are used to track the temperature evolving over time in the pipe, which are placed at [0.25,0], [2.75,0], [3.5,1.25], [4.25,0], [6.75,0].

The time-dependent material temperature for the five tracking probes is shown in Figure 5, compared with the reference results in \cite{xu2021positive}. The material temperature over time simulated by UGKP method is similar as the reference solution. In addition, Figure 6 (a)-(d) depict the material temperatures simulated by UGKP method at 20, 80, 150 and 300 ns, respectively. It is shown that the interface between the optically thin and thick materials is captured sharply by the UGKP method.
\begin{figure}
  \centering
  \includegraphics[width=0.6\textwidth]{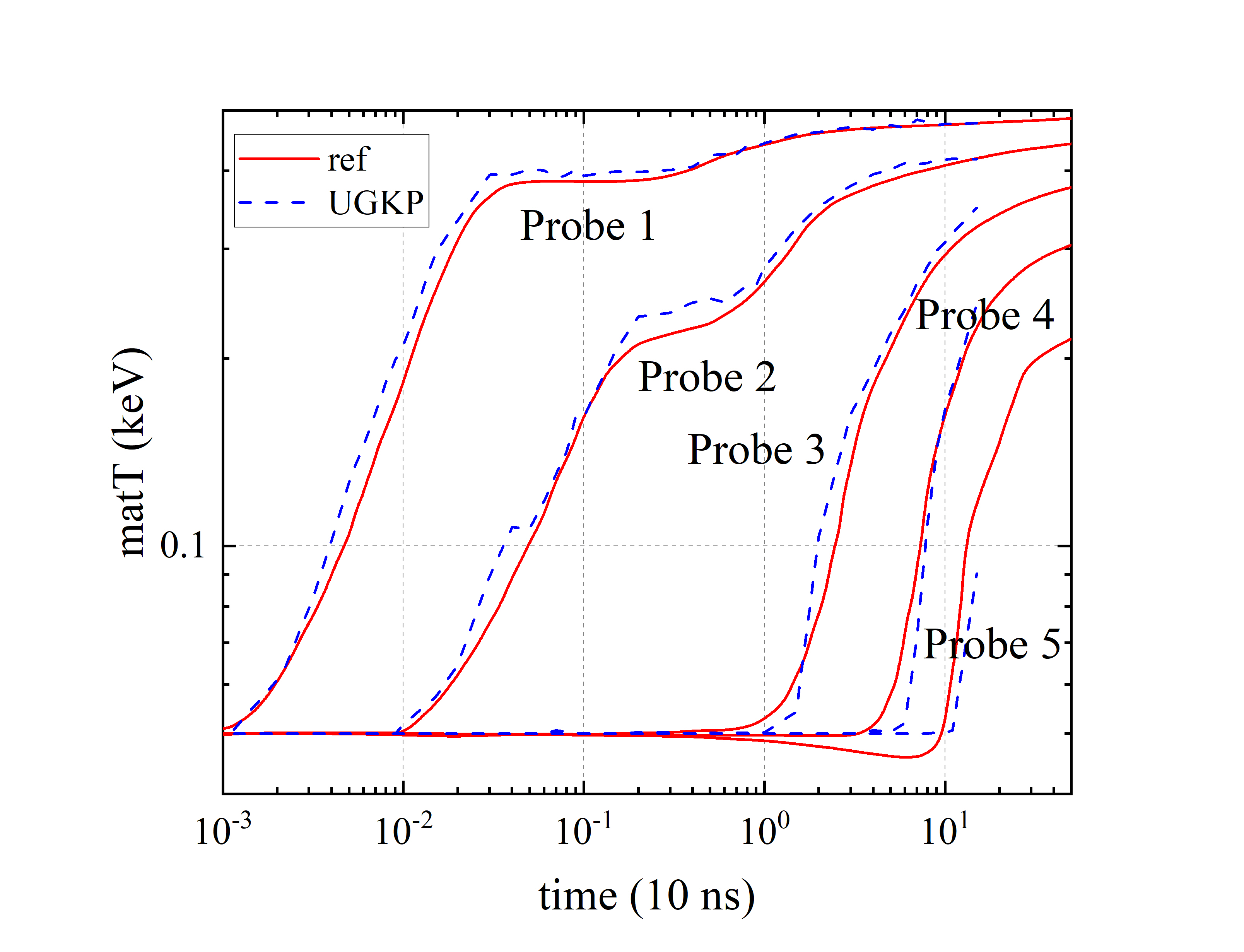}
  \caption{The material temperature over time for the five probes in the Tophat problem, compared with the reference solutions.}
  \label{fig_image7}
\end{figure}

\begin{figure}
  \centering
  \subfigure[]{\includegraphics[width=0.49\textwidth]{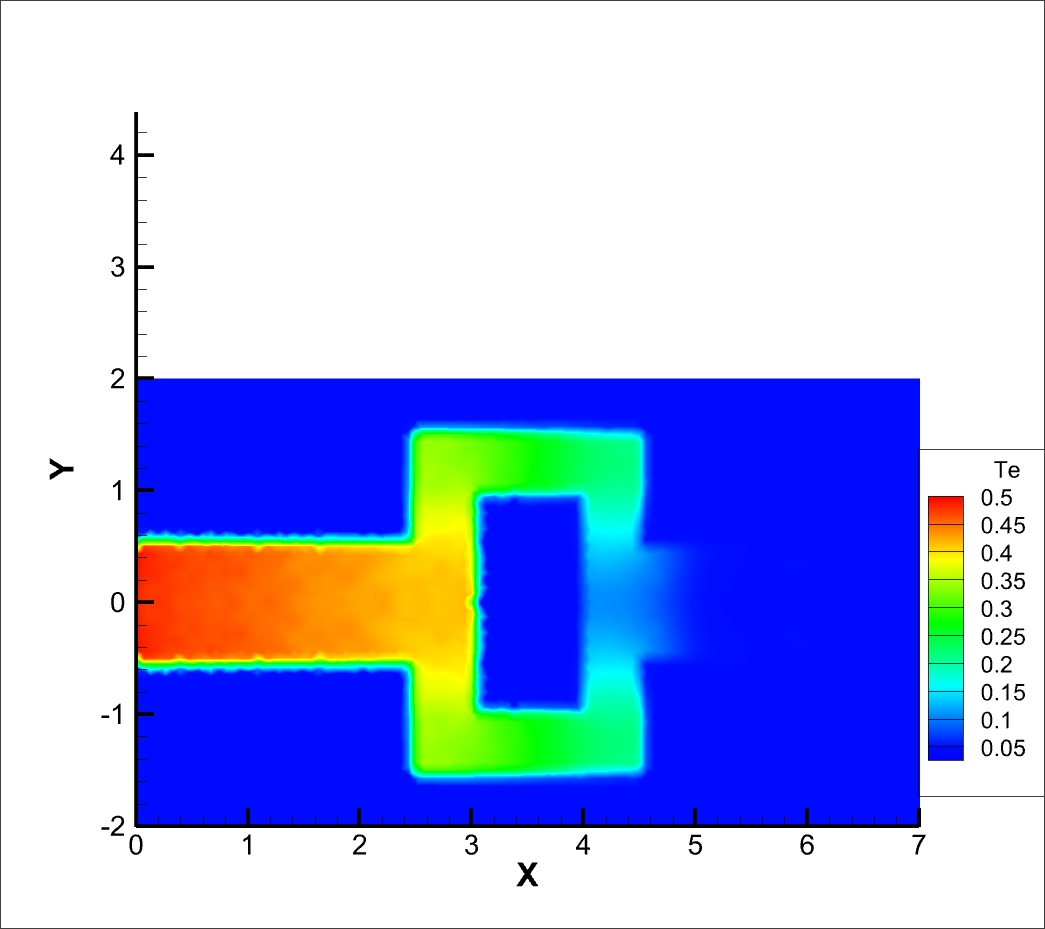}}
  \subfigure[]{\includegraphics[width=0.49\textwidth]{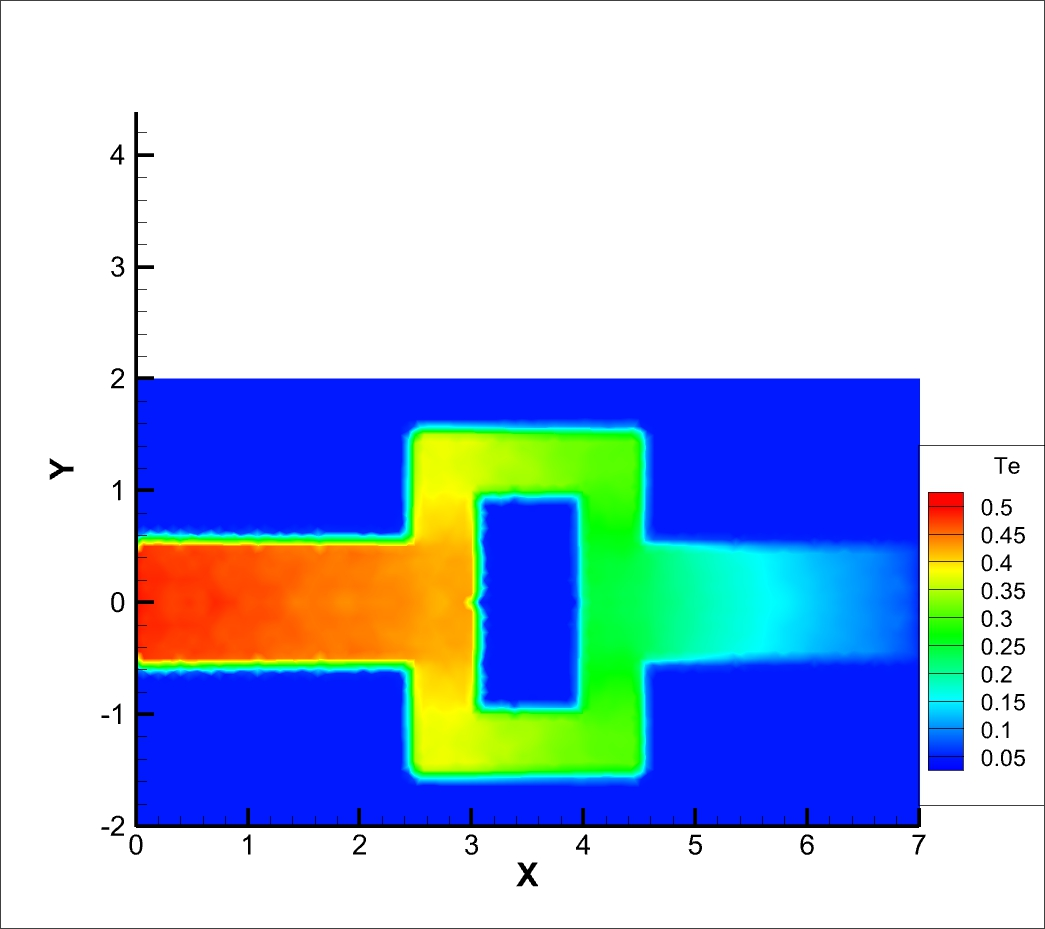}}
  \caption{The material temperatures simulated by UGKP method at time (a) 20 ns, (b) 80 ns, (c) 150 ns, (d) 300 ns.}
  \label{fig_image9}
\end{figure}

\subsection{5.4 Multi-group problem}
With the Marshak wave problems and the Tophat problems above, we have validated the extended UGKP method for the gray radiative transfer cases. The next example is a multi-group problem which we designed to test the extended UGKP method for the frequency-dependent radiative transfer cases. In this multi-group problem, we employ only 3 frequency groups for testing purpose. The layout of this multi-group problem is shown in Figure 7. The size of the computational domain is [0,0.2]$\mathrm{\times}$[-0.05,0.05], which consists of unstructured mesh with maximum size of 0.02 cm. In addition, it consists of 3 zones: one frequency-independent zone (zone A) and tow frequency-dependent zones (zone B and C), with detailed opacity and specific heat listed in Table 1. The computational domain is initially in equilibrium at 0.05 keV, and a 0.5 keV surface source with a Planck distribution is located on the left boundary. We also take eight probes to track the material temperature evolving over time in different zone at different places, which are placed at [-0.025,0.05], [0.025,0.05], [-0.025,0.1], [0.025,0.1], [-0.025,0.15], [0.025,0.15], [-0.025,0.2], [-0.025,0.2]. The location of these probes is also shown in Figure 7.

\begin{figure}
  \centering
  \includegraphics[width=0.6\textwidth]{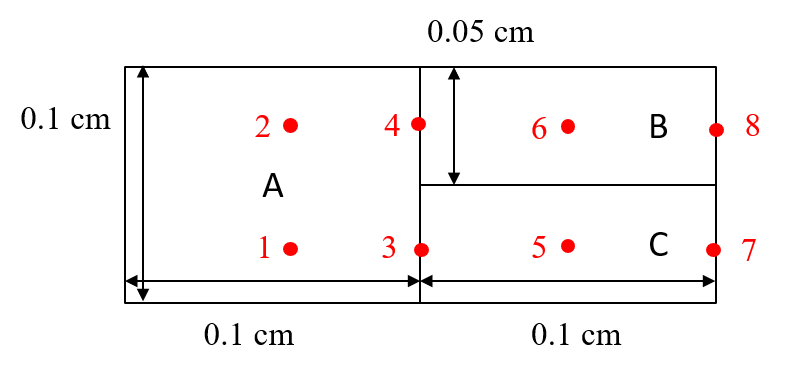}
  \caption{The layout of the multi-group problem.}
  \label{fig_image11}
\end{figure}
 
Table 1. The opacity and specific heat for the 3 zones in the multi-group problem.
\begin{tabular}{|p{0.8in}|p{0.8in}|p{0.8in}|p{0.8in}|p{0.8in}|} \hline
Zone & Specific heat [GJ/keV/cm${}^{3}$] & \multicolumn{3}{|p{2.3in}|}{Opacity [cm${}^{-1}$]} \\ \hline
 &  & Group 1 & Group 2 & Group 3 \\ \hline
A & 0.01 & 10 & 10 & 10 \\ \hline
B & 0.2 & 1000 & 100 & 10 \\ \hline
C & 0.2 & 100 & 10 & 1000 \\ \hline
\end{tabular}

To clearly show the difference between the radiation energy of different frequency groups transmitting over time, we compared the radiation energy at different places. First, the radiation energy of 3 frequency groups incident from left boundary source is shown in Figure 8 (a). It can be seen that the input radiation energy is dominated by frequency group 3. Figure 8 (b) and (c) give the radiation energy changing over time at probe 5 and probe 6, respectively. At the beginning, the radiation energy of both probe 5 and probe 6 is dominated by frequency group 1 and 2. With time increasing, the frequency group 2 becomes the leading group for probe 5, while the leading group changes into the frequency group 3 for probe 6. This difference is caused by the different opacities of frequency groups on zone B and C. For zone B, it has the smallest opacity for frequency group 3, which make the frequency group 3 to be the leading group for probe 6. However, the frequency group with the smallest opacity changes into the frequency group 2 for zone C, which also leads to the shift in the leading group. What's more, the system becomes stable with time further increasing. And the radiation energy of different frequency groups become similar as the input energy for both probe 5 and 6.

\begin{figure}
  \centering
  \subfigure[]{\includegraphics[width=0.49\textwidth]{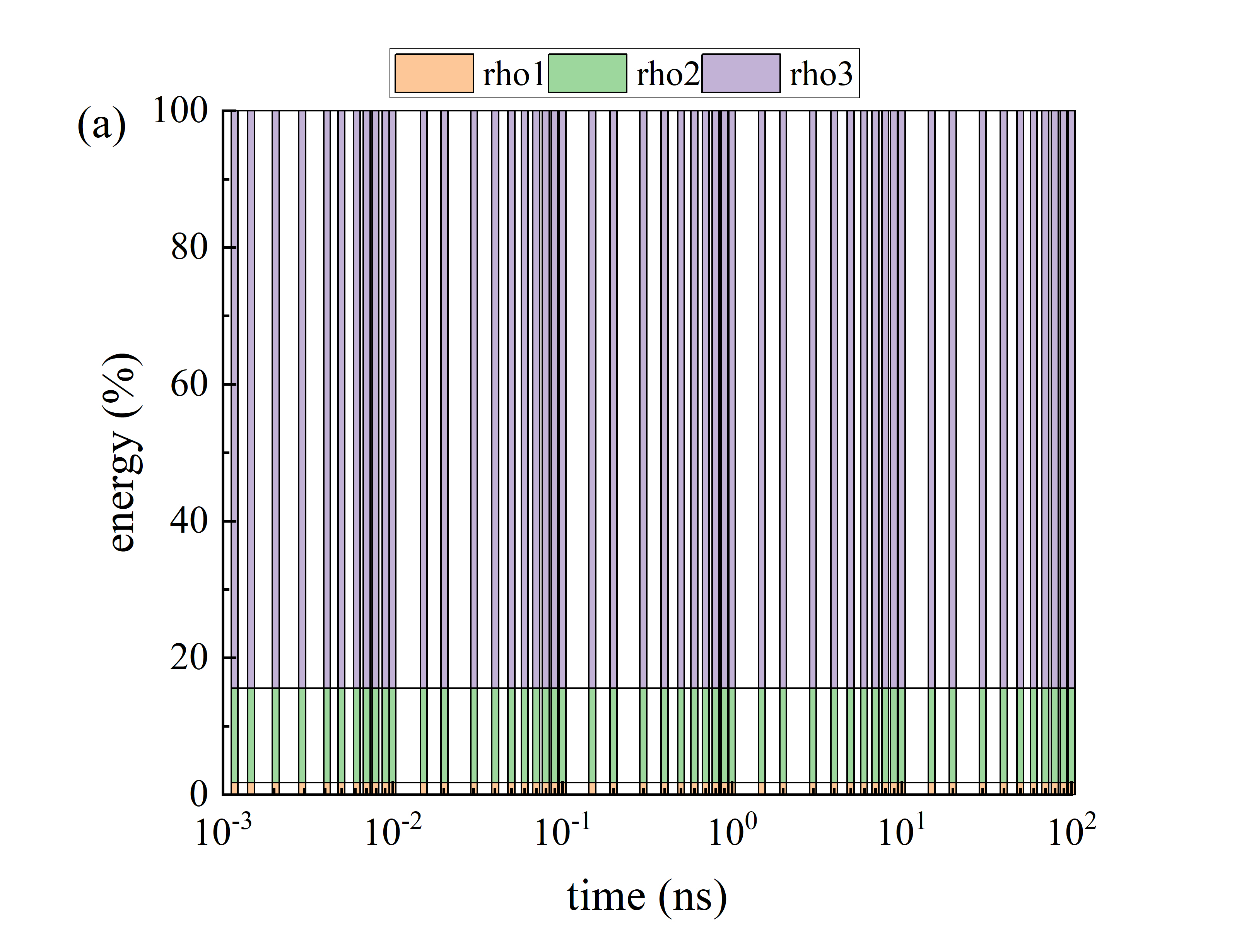}}
  \subfigure[]{\includegraphics[width=0.49\textwidth]{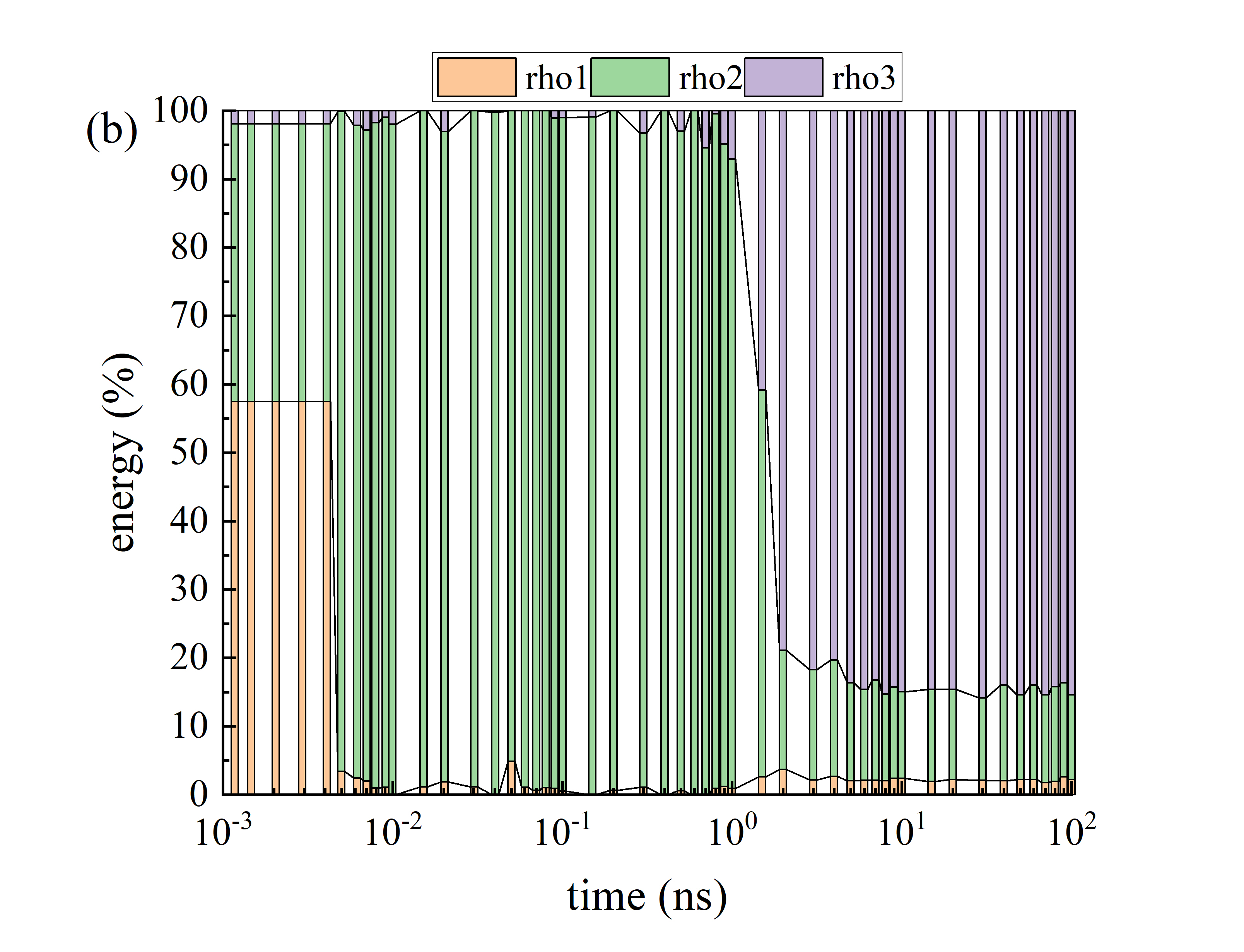}}
  \caption{The radiation energy of 3 different frequency groups over time for (a) the left boundary source, (b) probe 5, (c) probe 6 in the multi-group problem.}
  \label{fig_image12}
\end{figure}

We also compared the material temperature evolving over time at the eight probes above, which are shown in Figure 9. With frequency-independent opacity in zone A, the temperature is the same for probe 1 and 2. For probe 3, its material temperature is higher than probe 4, which is also caused by the different opacities of frequency groups on zone B and C. With the highest opacity for frequency group 3 on zone C, the input radiation energy (dominated frequency group 3) is more likely to be deposited on zone C. However, the smallest opacity for frequency group 3 on zone B makes the input radiation energy unlikely to be absorbed by the material. Thus, the material temperature of probe 4 is smaller than probe 3. In addition, the more input energy deposited on zone C, the fewer input energy transmits on zone C, which makes the radiation transport on zone C is slower than zone B. This can be clearly seen by the material temperatures of probe 5, 6, 7 and 8 in Figure 9. The material temperature of probe 5 is smaller than probe 6 during the radiation transport, and probe 7 and 8 have similar results. These results confirm the capability of the extended UGKP method for the frequency-dependent radiative transfer cases.

\begin{figure}
  \centering
  \includegraphics[width=0.6\textwidth]{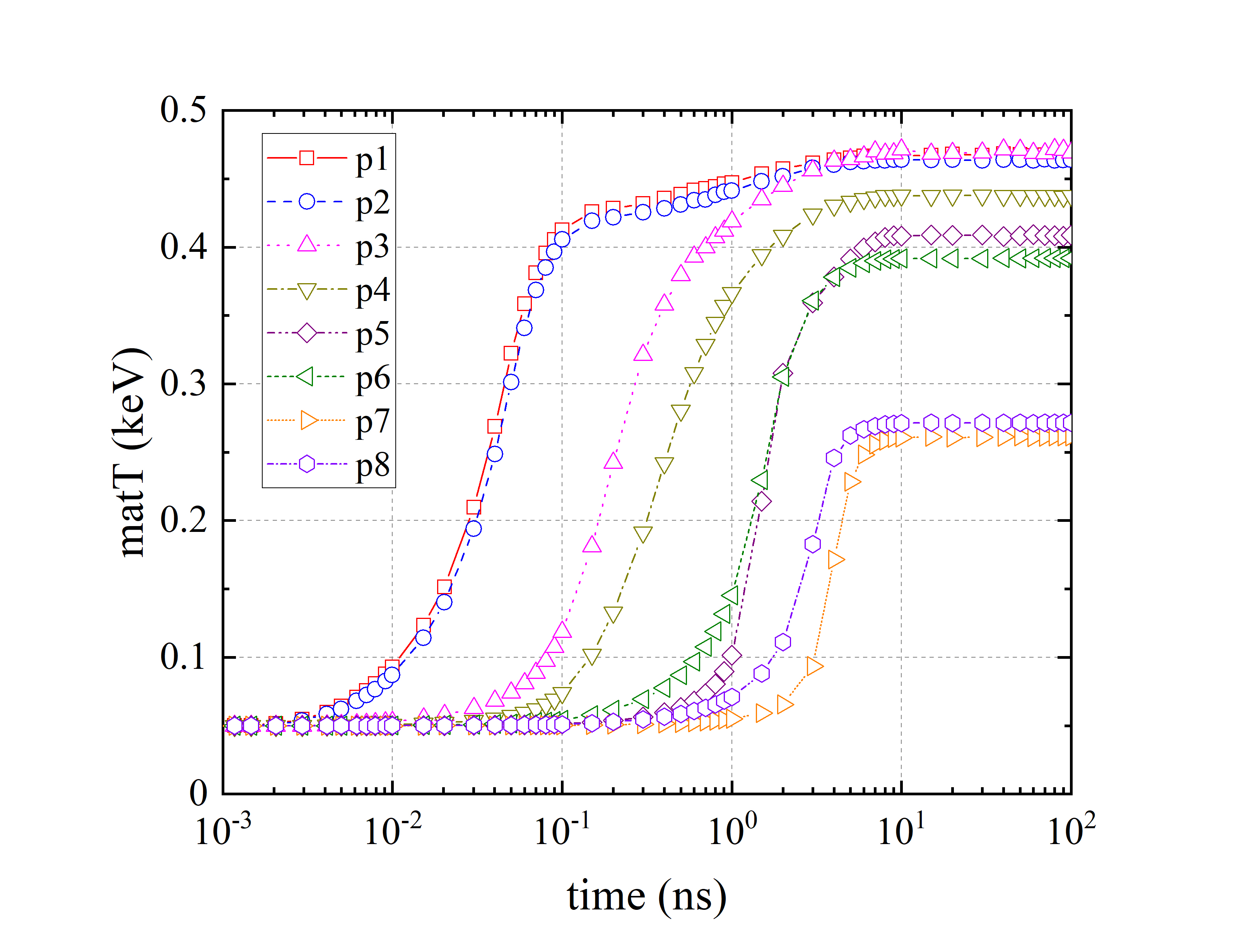}
  \caption{The material temperature over time for the eight probes in the multi-group problem.}
  \label{fig_image15}
\end{figure}

\subsection{Marshak wave-2A-scatter problem}
To further test the extended UGKP method for scattering process considered cases, we modified the Marshak wave problems. The modified Marshak wave-2A-scatter problem is exactly the same as the Marhsak wave-2A problem, except that a temperature-dependent scattering coefficient with $\sigma ={30.0\mathord{\left/ {\vphantom {30.0 T^{3} }} \right. \kern-\nulldelimiterspace} T^{3} } cm^{-1} $ is also considered. With this temperature-dependent scattering coefficient, the total opacity in this problem is twice the Marhsak wave-2A problem.

The material temperature from UGKP simulation and the diffusion equation solution for this problem is given in Figure 10 (a). As we can see, the difference between the UGKP simulation and the diffusion equation solution in this problem is much smaller than the Marshak wave-2A problem in Figure 3 (b), which is due to the larger total cross-sections. We also compared the material temperature from the diffusion equation solution in this problem with the Marshak wave-2A problem, which is shown in Figure 10 (b). With twice total opacity in this problem, the diffusion coefficient will be a half as in the Marshak wave-2A problem. Thus, the radiation propagation speed is also a half as in the Marshak wave-2A problem. This is clearly shown in Figure 10 (b) that the results at times 0.4 and 0.8 ns for this problem is similar as the results at times 0.2 and 0.4 ns for the Marshak wave-2A problem. The material temperature from the UGKP simulation in this problem is also compared with the Marshak wave-2A problem in Figure 10 (c), which also shows the influence of the scattering process considered in this problem.

\begin{figure}
  \centering
  \subfigure[]{\includegraphics[width=0.32\textwidth]{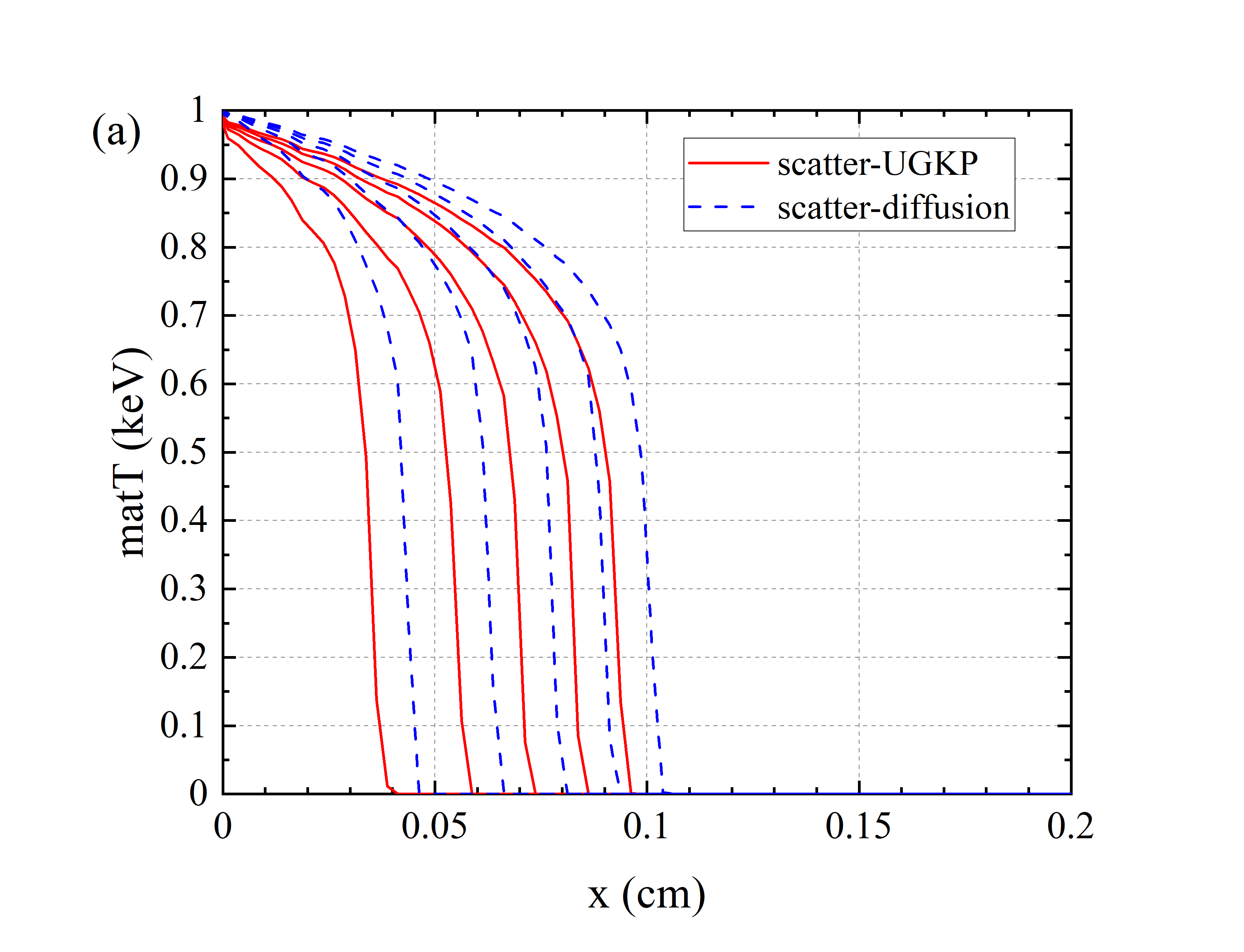}}
  \subfigure[]{\includegraphics[width=0.32\textwidth]{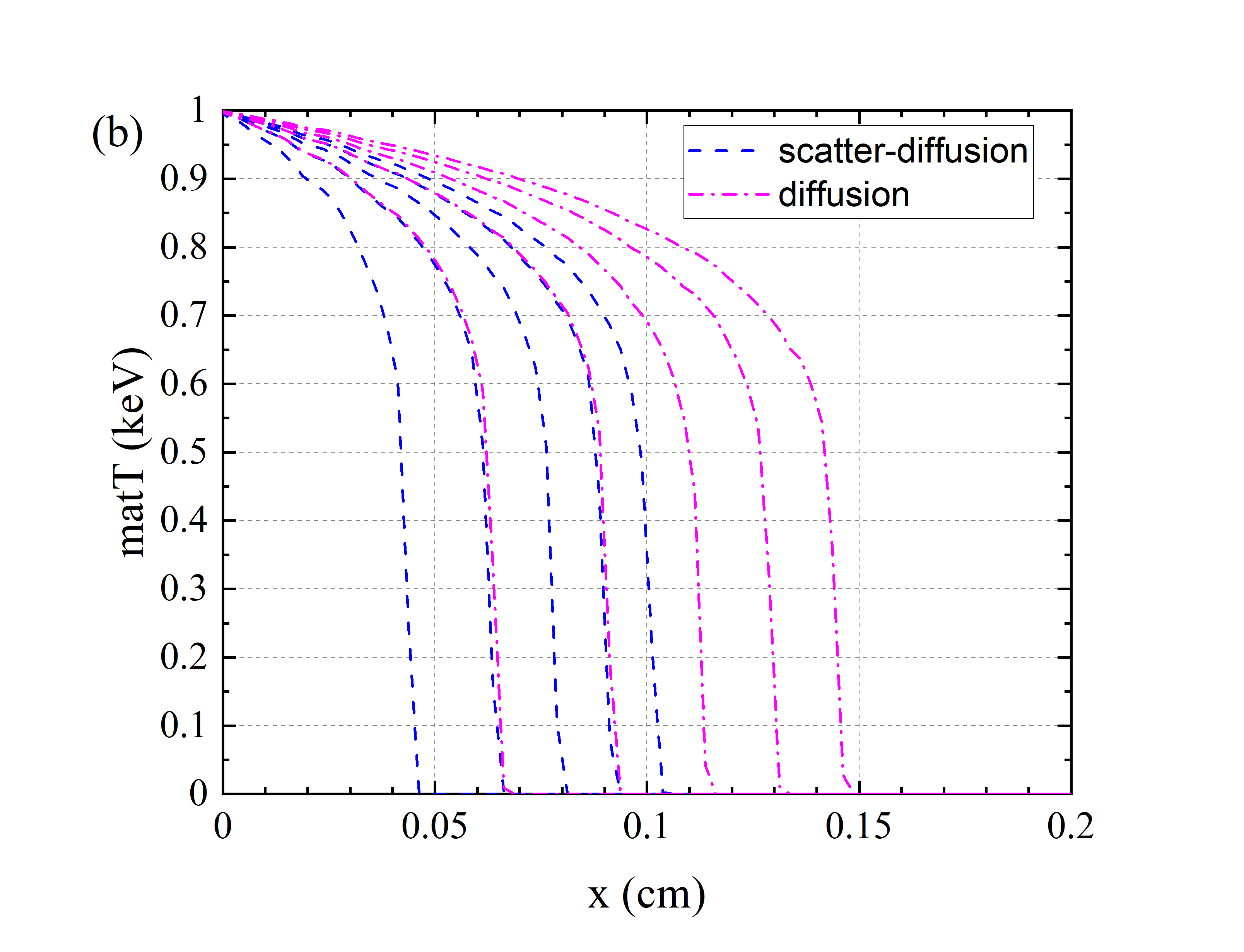}}
  \subfigure[]{\includegraphics[width=0.32\textwidth]{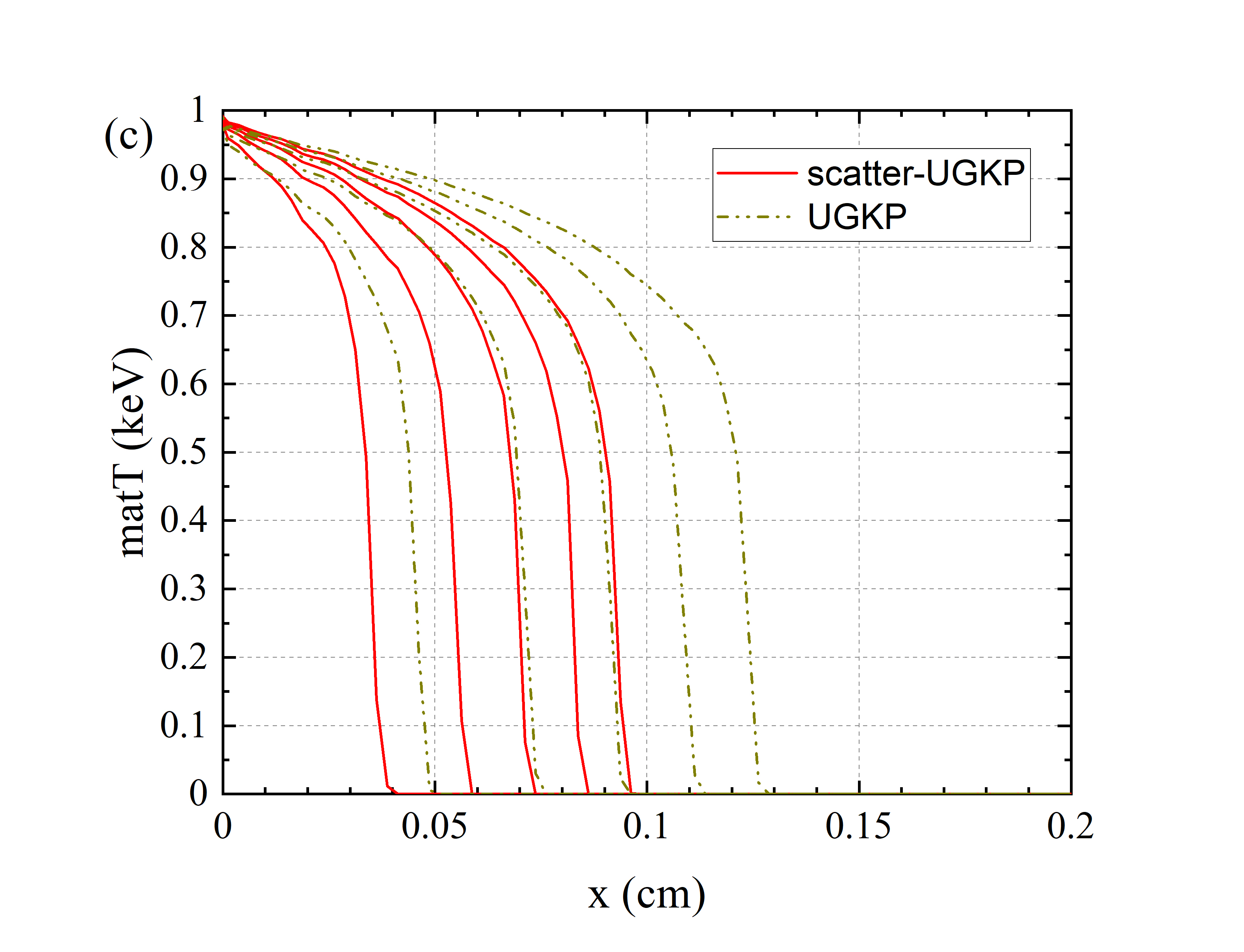}}
  \caption{The material temperatures at times 0.2, 0.4, 0.6, 0.8, 1.0 ns respectively. (a) The material temperature from UGKP simulation and the diffusion equation solution. (b) The material temperature from the diffusion equation solution compared with the Marshak wave-2A problem. (c) The material temperature from the UGKP simulation compared with the Marshak wave-2A problem.}
  \label{fig_image16}
\end{figure}

\subsection{5.6 Marshak wave-2B-scatter problem}
The modified Marshak wave-2B-scatter problem is also exactly the same as the Marhsak wave-2B problem, except that a temperature-dependent scattering coefficient with $\sigma ={30{\rm 0}.0\mathord{\left/ {\vphantom {30{\rm 0}.0 T^{3} }} \right. \kern-\nulldelimiterspace} T^{3} } cm^{-1} $ is considered. This also makes the total opacity in this problem be twice as the Marhsak wave-2B problem.

Figure 11 (a) shows the material temperature from UGKP simulation and the diffusion equation solution for this problem. Since the absorption/emission coefficient in Marshak wave-2B problem is large enough to get the equilibrium diffusion limit solution, twice the total opacity in this problem will also get the equilibrium diffusion limit solution. Thus, the UGKP results are also close to the diffusive limit results. In addition, the radiation propagation speed for both UGKP simulation and the diffusion equation solution is also a half as in the Marshak wave-2B problem, due to the twice the total opacity in this problem. This is shown in Figure 11 (b) and (c) that the results at times 30 and 60 ns for this problem is similar as the results at times 15 and 30 ns for the Marshak wave-2B problem.

\begin{figure}
  \centering
  \subfigure[]{\includegraphics[width=0.32\textwidth]{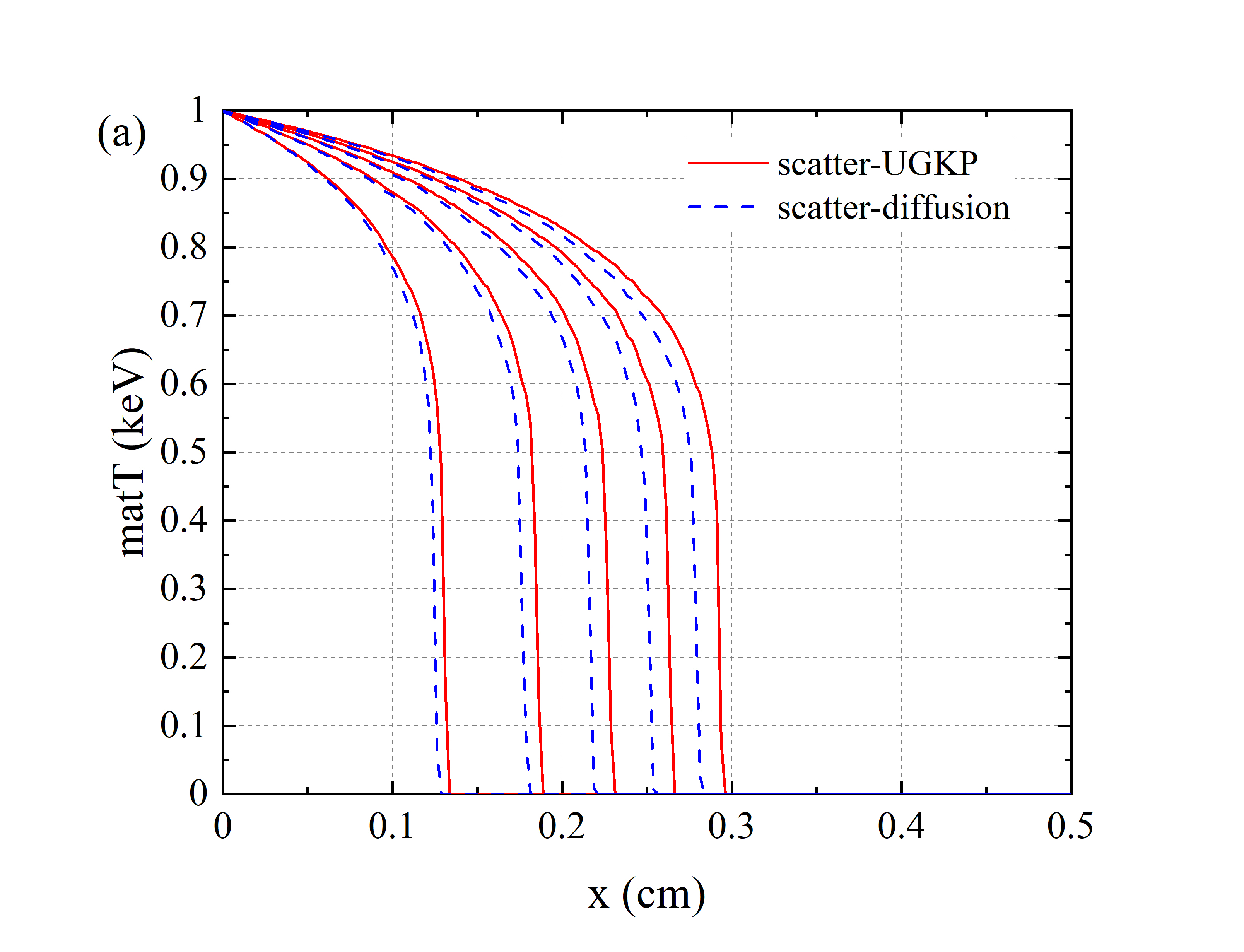}}
  \subfigure[]{\includegraphics[width=0.32\textwidth]{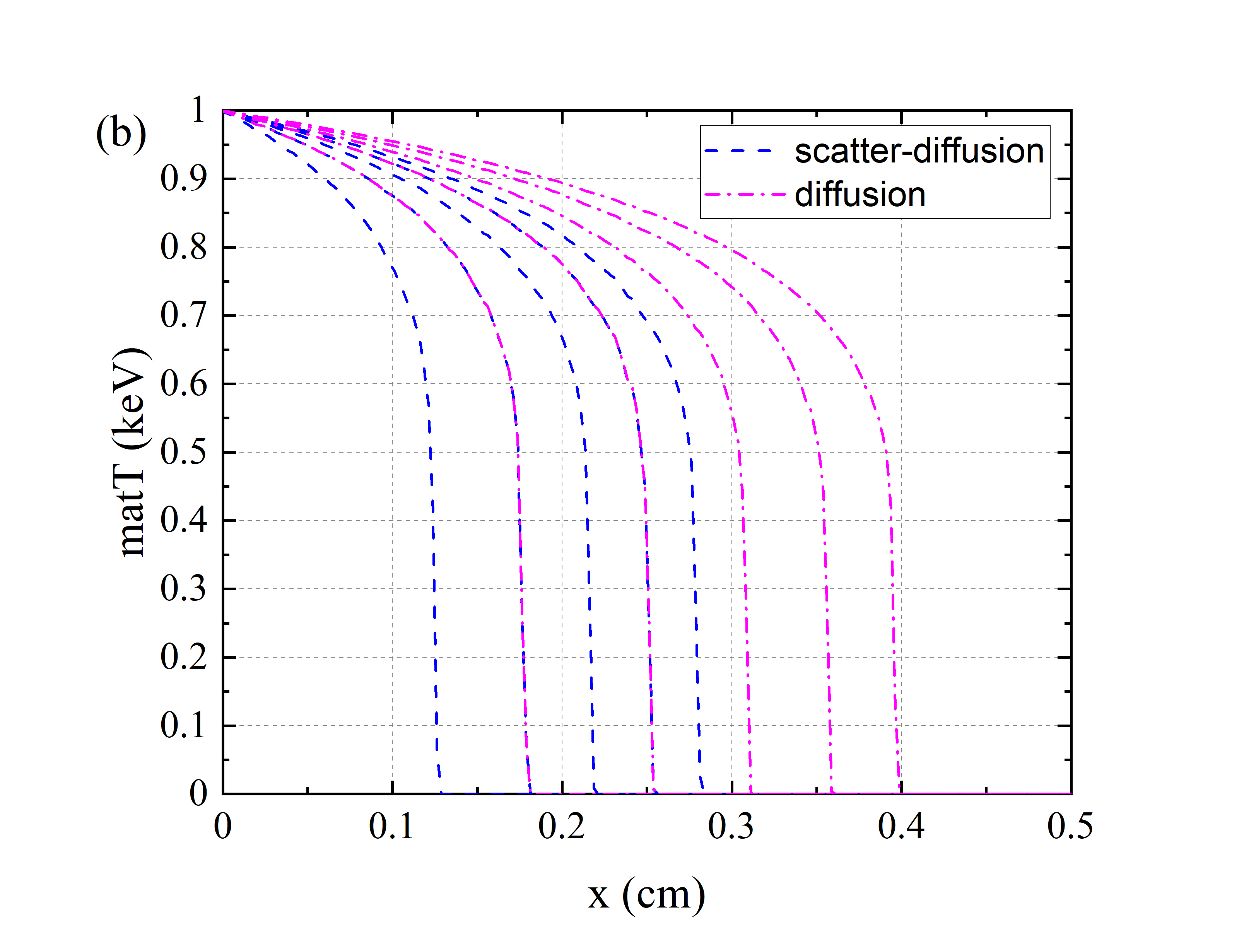}}
  \subfigure[]{\includegraphics[width=0.32\textwidth]{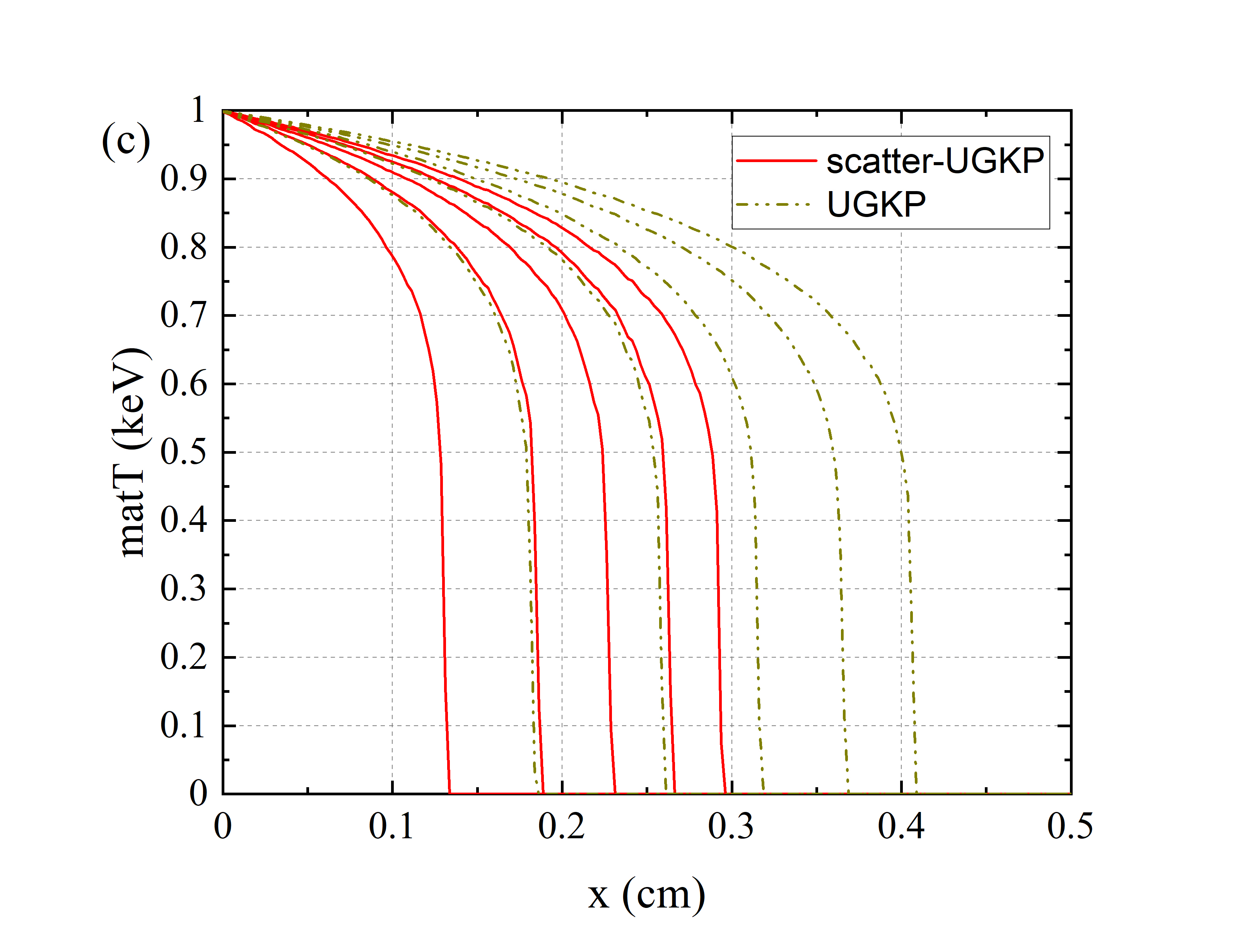}}
  \caption{The material temperatures at times 15, 30, 45, 60, 74 ns respectively. (a) The material temperature from UGKP simulation and the diffusion equation solution. (b) The material temperature from the diffusion equation solution compared with the Marshak wave-2B problem. (c) The material temperature from the UGKP simulation compared with the Marshak wave-2B problem.}
  \label{fig_image19}
\end{figure}

\section{Conclusions}\label{conclusions}
In this work, the UGKP method has been extended to solve the frequency-dependent radiative system with both absorption/emission and scattering process under unstructured mesh. The extended UGKP method has the asymptotic preserving (AP) property in both optical thin and thick regime for a continuum spectrum in the frequency domain. In addition, a smooth transition in the regime between the diffusion and free transport limit could be achieved by the extended UGKP method. Some numerical simulation results are presented to show the capability of the extended UGKP method. The results of the one-dimensional Marshak wave problems and the two-dimensional Tophat problem confirm that this extended UGKP method could degenerate into the gray radiative transfer cases. What's more, the results of a self-designed multi-group problem validate the extended UGKP method for the frequency-dependent radiation cases. Finally, the Marshak wave problems have also been modified to test the extended UGKP method. These results show the capacity of the extended UGKP method. In the future, we will extend it to the system with anisotropic scattering process.

\section*{Credit authorship contribution statement}
Yuan Hu: Methodology, Software, Investigation, Writing -- original draft.
Chang Liu: Methodology, Software, Writing -- review \& editing, Supervision, Resources.

\section*{Acknowledgement}
We thank Dr. Yanli Wang from Beijing Computational Science Research Center for helpful discussions and providing computational resources. 
Chang Liu is partially supported by the National Natural Science Foundation of China (12102061), 
the Presidential Foundation of the China Academy of Engineering Physics (YZJJZQ2022017),
and the National Key R\&D Program of China (2022YFA1004500).

\bibliographystyle{unsrt}
\bibliography{UGKP}
\end{document}